\documentclass[12pt]{article}
\pdfoutput =1
\usepackage{graphicx}
\usepackage{float} 
\usepackage{subfloat}
\usepackage{amsmath}
\usepackage{slashed}
\usepackage{amsfonts}   
\usepackage{amssymb}
\usepackage[colorlinks]{hyperref}
\hypersetup{
  colorlinks,
  citecolor=blue,
  linkcolor=red,
  urlcolor=violet}
\usepackage{enumerate}
\numberwithin{equation}{section}
\usepackage{subcaption}

\begin{document}
\date{}
\title{{\bf{\Large Holographic JT gravity with quartic couplings}}}
\author{

 {\bf {\normalsize Hemant Rathi\thanks{E-mail:  hrathi07@gmail.com, hrathi@ph.iitr.ac.in}~ and ~Dibakar Roychowdhury}$
$\thanks{E-mail:  dibakarphys@gmail.com, dibakar.roychowdhury@ph.iitr.ac.in}}\\
 {\normalsize  Department of Physics, Indian Institute of Technology Roorkee,}\\
  {\normalsize Roorkee 247667, Uttarakhand, India}
\\[0.3cm]
}
\maketitle
\begin{abstract}
 We construct the most general theory of 2D Einstein-dilaton gravity coupled with $U(1)$ gauge fields that contains all the 2-derivative and the 4-derivative interactions allowed by the diffeomorphism invariance. We renormalise the 2D action and obtain the vacuum solution as well as the black hole solution. The vacuum solution in the UV is dominated by Lifshitz$_2$ with dynamical exponent ($z=\frac{7}{3}$) while on the other hand, the spacetime curvature diverges as we move towards the deep IR limit. We calculate the holographic stress tensor and the central charge for the  boundary theory. Our analysis shows that the central charge goes as the inverse power of the coupling associated to 4-derivative interactions. We also compute the Wald entropy for 2D black holes and interpret its near horizon divergence in terms of the density of states. We compare the Wald entropy with the Cardy formula and obtain the eigen value of Virasoro operator ($L_0$) for our model. Finally, we explore the near horizon structure of 2D black holes and calculate the central charge corresponding to the CFT near horizon. We further show that the near horizon CFT may be recast
as a 2D Liouville theory with higher derivative corrections.  We study the Weyl invariance of this generalised Liouville theory and identify the Weyl anomaly associated to it.  We also comment on the classical vacuum structure of the theory. 
 \end{abstract}
\section{Introduction}
It has been more than two decades now since the discovery of the celebrated AdS/CFT correspondence \cite{Maldacena:1997re}-\cite{ Gubser:1998bc}. The $AdS_{d+1}/CFT_d$ correspondence (also known as gauge/gravity duality) claims an equivalence between the partition functions of strongly coupled gauge theories in d-dimensions and the  classical gravitational counterpart living in d+1 dimensions. In other words, the duality offers a suitable platform  to examine the strongly coupled systems using weakly coupled dual gravitational descriptions. 

In principle, this duality holds for any d dimensional space-time. However, the study of this duality in $d=1$ dimension provides a remarkable insight about models of quantum gravity living in two dimensions \cite{Strominger:1998yg}-\cite{Azeyanagi:2007bj}.

 Following the spirit of the above discussions, the present paper focuses on a particular model of classical gravity in two dimensions known as the Jackiw-Teitelboim (JT) gravity \cite{Jackiw:1984je,Teitelboim:1983ux}. It is the 2D version of Einstein's gravity (with negative cosmological constant) coupled to a dilatonic field. 
 
 The holographic dual of JT gravity is conjectured to be the Sachdev-Ye-Kitaev (SYK) model \cite{Sachdev:1992fk}-\cite{Gross:2016kjj} which describes the quartic interactions among the Majorana fermions. The striking feature of the SYK model is that it is exactly solvable in the limit of strong coupling and in the large N limit.
 
 In literature, the duality between the SYK model and JT gravity has been explored in the presence of U(1) gauge fields \cite{Gaikwad:2018dfc}-\cite{Davison:2016ngz} as well as the SU(2) Yang-Mills fields \cite{Lala:2020lge}. In particular, the authors in \cite{Castro:2008ms} consider the JT gravity model minimally coupled to the U(1) gauge fields and study the holographic stress tensor and the central charge \cite{Hartman:2008dq,Cadoni:2000gm,Castro:2008ms}, \cite{deHaro:2000wj}-\cite{Hohm:2010jc} for the boundary theory.

The purpose of the present article is to look for a further generalisation on the gravitational side of the correspondence by incorporating higher derivative (quartic) interactions in the same spirit as that of its cousins living in higher dimensions\footnote{It was observed in \cite{Utiyama:1962sn} that the 4-derivative interactions are crucial to obtain the finite result of average stress tensor of quantum fields coupled with classical gravity. Higher derivative corrections are also found to be useful in cosmology in order to describe the inflationary models, see \cite{Castellanos:2018dub}-for a recent review.}\cite{Lovelock:1971yv}-\cite{Kraus:2005vz} .
  
In the present paper, we start with the most generic higher derivative theories of gravity in five dimensions \cite{Myers:2009ij} and search for its imprints in the lower dimensional models in the context of SYK/JT gravity correspondence. We follow the standard procedure of dimensional reduction \cite{Davison:2016ngz} which results in the most generic higher derivative theories of gravity (including quartic interactions) in 2D. 

Following the standard AdS/CFT prescription, we compute various physical observables associated  with the dual quantum mechanical model in 1D and explore upon the effects of incorporating the higher derivative corrections on the dual field theory observables. In particular, we compute the holographic stress energy tensor and estimate the central charge associated with the 1D boundary theory.

We also construct the corresponding 2D black hole solutions and obtain the central charge using Cardy formula \cite{Cadoni:2000gm}, \cite{Cardy:1986ie}-\cite{Strominger:1997eq}. Finally, we show that the model of JT gravity with quartic coupling can be recast as a ``generalised'' 2D Liouville theory \cite{Solodukhin:1998tc}-\cite{Mertens:2020hbs} of quantum gravity with some complicated potential function.
  
The organisation and the summary of results of the paper is as follows :

$\bullet$ In Section (\ref{S2}), following the standard procedure of dimensional reduction, we construct the most general theory of 2D Einstein-dilaton gravity coupled with U(1) gauge fields. Our theory contains all possible 2-derivative as well as the 4-derivative interaction terms allowed by the diffeomorphism invariance. 

$\bullet$ In Section (\ref{S3}), we obtain vacuum solutions of the 2D theory by treating the higher derivative interactions as ``perturbations''. We observe that the scalar curvature corresponding to the 2D theory diverges in the deep IR limit due to the presence of higher derivative interactions. On the other hand, the vacuum solution in the UV limit is dominated by Lifshitz$_2$ with dynamical exponent ($z=\frac{7}{3}$). On the other hand, if we switch off the 4-derivative interactions then, the space-time geometry becomes $AdS_2$ in IR limit and Lifshitz$_2$ (with dynamical exponent $z=\frac{3}{2}$) in UV limit which is consistent with  \cite{Lala:2020lge} . 

$\bullet$ In Section (\ref{S4}), we obtain the Gibbons-Hawking-York (GHY) boundary terms \cite{Gibbons:1976ue}-\cite{York:1972sj} for the 2D model that is needed for the successful implementation of the variational principle. Finally, we estimate counter terms which lead to the ``renormalised'' action.

$\bullet$ In Section (\ref{S5}), we use the renormalised action  to determine the boundary stress tensor and the central charge in the Fefferman Graham gauge \cite{fefferman}. We observe that the central charge associated with the boundary theory goes as the inverse power of the quartic coupling ($\kappa$). This further implies that, a smooth $\kappa\rightarrow0$ limit of the central charge does not exist.

$\bullet$ In Section (\ref{SBH}), we obtain black hole solution for the 2D theory by treating the higher derivative interactions as a perturbative corrections over the pure JT gravity solutions.

$\bullet$ In Section (\ref{SBH1}), we explore thermal properties of 2D black holes in our model. In particular, we discuss the Wald entropy \cite{Wald:1993nt}-\cite{Pedraza:2021cvx} for 2D black holes and observe that the Wald entropy diverges near the horizon due to the presence of higher derivative interactions. We interpret these divergences in terms of the density of states \cite{thooft}-\cite{Brustein:2010ms}. Finally we compare the Wald entropy with the Cardy formula for 2D black holes and estimate the eigen value of the Virasoro operator ($L_0$) for our model. 

$\bullet$ In Section (\ref{SS8}), we investigate the near horizon structure of 2D black holes in the presence of quartic interactions. We observe that the trace of the stress tensor vanishes in the near horizon limit which indicates the presence of a conformal field theory in the vicinity of the horizon. Finally, we transform the 2D theory into the ``generalised'' Liouville theory \cite{Solodukhin:1998tc}-\cite{Mertens:2020hbs} using the proper field re-definition and calculate the associated central charge. We observe that the central charge corresponding to the generalised Liouville theory diverges due to the presence of higher derivative interactions. 

$\bullet$ In Section (\ref{GLT}), we discuss the Weyl transformation \cite{Jackiw:2005su} properties of the generalised Liouville theory. We observe that the generalised Liouville theory is not invariant under the Weyl re-scaling. On top of it, the trace of its stress tensor does not vanish and comes out to be proportional to its central charge. We identify this as the Weyl (or trace) anomaly \cite{Castro:2019vog} for the generalised Liouville theory.

$\bullet$ Finally, in Section (\ref{sum}), we draw our conclusion with some future remarks.

\section{Construction of the \texorpdfstring{$2D$}{2D} action }\label{S2}
The purpose of this section is to discuss the basic methodology that leads to the most general 2D action for Einstein-dilaton gravity coupled to U(1) gauge fields. We start with the most general theory of Einstein gravity coupled with U(1) gauge fields in five-dimensions \footnote{See Appendix \ref{Non-abelian} for a discussion on the Non-abelian sector.} \cite{Myers:2009ij}.
\begin{align}\label{action 5D}
    S_{(5D)}=&\int d^5x\sqrt{-g_{(5)}}\Big [(12+R)-\frac{\eta_1}{4}F^2+\eta_2[R_{MNOP}]^2+\eta_3 F^4+\eta_4F^{SP}F_{PR}F^{RQ}F_{QS}+ \nonumber\\
&\eta_5\bigtriangledown_MF^{MN}\bigtriangledown^OF_{ON}+\epsilon^{MNOPQ}\Big(\eta_{6}F_{MN}F_{OP}\bigtriangledown^RF_{RQ}+\eta_{7}F_{MN}F_{OR}\bigtriangledown^RF_{PQ}+\nonumber\\
&\eta_{8}F_{MN}F_{OR}\bigtriangledown_P{F_Q}^R+\eta_{9}A_MR_{NOIJ}{R_{PQ}}^{IJ}\Big) \Big ]
\end{align}
where, $\eta_i$ $(i=1,..,9)$ are the respective coupling constants. The key feature of this model is that it contains the 4-derivative interaction terms along with the usual 2-derivative interactions. These higher derivative terms are the key contents of our model.

In principle, it is possible to add several other 4-derivative terms to the above action (\ref{action 5D}). However, all such terms can be eliminated using a proper redefinition of fields as demonstrated in Appendix \ref{gen action 5D}. Therefore, the action (\ref{action 5D}) is the most general theory of gravity (coupled to U(1) gauge fields) containing both 2-derivative and 4-derivative interaction terms.

We are interested in studying the JT gravity model with chemical potential in the context of $AdS_2/CFT_1$ correspondence. On that note, we will require to get rid of the extra dimensions present in the 5D theory (\ref{action 5D}).

Systematically, this can be achieved following a reduction ansatz for the metric as well as the gauge field \cite{Davison:2016ngz}
\begin{align}\label{ansatz}
    ds^2_{(5)}&=ds^2_{(2)}+\phi(t,z)^{\frac{2}{3}}(dx^2+dy^2+dz^2),\hspace{2mm}A_Mdx^M=A_{\mu}dx^{\mu},\hspace{2mm} A_{\mu}\equiv A_{\mu}(x^{\nu})
\end{align}
where M is the 5 dimensional index and $\mu$ stands for the 2 dimensional space-time index.

Using the above ansatz (\ref{ansatz}), one arrives at the required Einstein Hilbert action in 2 dimensions\footnote{See Appendix \ref{covariance} for a detailed discussion on the general covariance of the action.}
\begin{align}\label{action 2D}
S_{EH} =& \int d^2x\sqrt{-g_{(2)}}\phi \Bigg [(12+R)-\frac{\xi}{4}F^2+\kappa\Big[(R_{\mu\nu\alpha\beta})^2+\frac{3}{4}\Big(\bigtriangledown_{\mu}\phi^{\frac{2}{3}}\Big)^4\nonumber+\\
&4\Big\{\frac{3}{4}\Big(\bigtriangledown_{\lambda}\phi^{\frac{2}{3}}\Big)\Big(\bigtriangledown_{\beta}\phi^{\frac{2}{3}}\Big)\Gamma^{\lambda}_{\alpha\mu}\Gamma^{\beta}_{\rho\sigma}g^{\alpha\rho}g^{\mu\sigma}+\frac{2}{3}\Gamma^{\lambda}_{\alpha\mu}\Big(\bigtriangledown_{\lambda}\phi^{\frac{2}{3}}\Big)(\bigtriangledown^{\alpha}\phi)(\bigtriangledown^{\mu}\phi)\phi^{\frac{-4}{3}}\nonumber\\
&-\Gamma^{\lambda}_{\alpha\mu}\Big(\bigtriangledown_{\lambda}\phi^{\frac{2}{3}}\Big)\{\partial_{\beta}(\bigtriangledown_{\sigma}\phi)\}g^{\alpha\beta}g^{\mu\sigma}\phi^{\frac{-1}{3}}-\frac{4}{9}\phi^{\frac{-5}{3}}(\bigtriangledown^{\alpha}\phi)(\bigtriangledown^{\mu}\phi)\{\partial_{\alpha}(\bigtriangledown_{\mu}\phi)\}+\nonumber\\
&\frac{4}{27}(\bigtriangledown_{\mu}\phi)^4\phi^{\frac{-8}{3}}-\frac{1}{3}\{\partial_{\alpha}(\bigtriangledown_{\mu}\phi)\}\{\partial_{\beta}(\bigtriangledown_{\rho}\phi)\}g^{\alpha\beta}g^{\mu\rho}\phi^{\frac{-2}{3}}\Big\}+F^4+F^{\mu\nu}F_{\nu\lambda}F^{\lambda\sigma}F_{\sigma\mu}+\nonumber\\
&\bigtriangledown_{\mu}F^{\mu\nu}\bigtriangledown^{\lambda}F_{\lambda\nu}\Big] \Bigg ].
\end{align}

Notice that, in order to arrive (\ref{action 2D}), we make a special choice of coupling constants namely, $\eta_1=\xi$ and $\eta_2=\eta_3=\eta_4=\eta_5=\kappa $. Furthermore we treat these coupling constants to be small enough such that the 2-derivative and 4-derivative interaction terms can be treated as pertubations over pure JT gravity. On variation of (\ref{action 2D}) one arrives at the following structure  
\begin{eqnarray}
\delta S_{EH} &=&\int d^2x\sqrt{-g_{(2)}}[H_{\mu\nu}\delta g^{\mu\nu}+H_{\phi}\delta\phi+H^{\mu}\delta A_{\mu}].
\end{eqnarray}

Equations of motion for the metric, dilaton and the gauge field in bulk will be given by equating  the individual coefficients $\sqrt{-g}H_{\mu\nu}$, $\sqrt{-g}H_{\phi}$ and $\sqrt{-g}H_{\mu}$ to zero. Technically speaking, it will be easy to handle these equations using the static gauge given below
\begin{eqnarray}\label{sg}
ds^2= e^{2\omega(z)}(-dt^2+dz^2), \hspace{3mm}A_{\mu}=(A_{t}(z),0).
\end{eqnarray}
\section{Vacuum solutions}\label{S3}
Even in the static gauge, it is difficult to solve the bulk equations of motion exactly. Therefore, we will prefer to solve these equations perturbatively treating $\xi$ and $\kappa$ as an expansion parameter. 

Systematically, one can expand these fields in terms of the expansion parameters as shown in equation (\ref{e1})-(\ref{e3})
\begin{eqnarray}
 \omega&=&\omega_{(0)}+\xi\omega_{(1)}+\kappa\omega_{(2)},\label{e1}\\
\phi&=&\phi_{(0)}+\xi\phi_{(1)}+\kappa\phi_{(2)},\label{e2}\\
A_{t} &=& A_{t(0)}+\frac{\kappa}{\xi} A_{t(1)}\hspace{1mm},\hspace{3mm}\Big|\frac{\kappa}{\xi}\Big|<<1.\label{e3}
\end{eqnarray}

 In the above equation, the subscript (0) in ($\phi$, $\omega$) denotes the pure JT gravity fields whereas subscripts (1) and (2) denote the contributions coming from 2-derivative and 4-derivative interaction terms in (\ref{action 2D}). Notice that, the expansion of the gauge field ($A_t$) is different from $\phi$ and $\omega$ because it is absent in pure JT gravity theory. Gauge fields start appearing in the action as 2-derivative and 4-derivative interaction with coupling constants $\xi$ and $\kappa$ respectively. Therefore, the subscripts (0) and (1) in $A_t$ denote the contributions due to 2-derivative and 4-derivative interaction terms respectively. Finally, using equation (\ref{e1})-(\ref{e3}) we expand the coefficients $H_{\phi}$ , $H_{\mu}$ and $H_{\mu\nu}$ as follows
\begin{align}\label{ee}
H_{\phi}=H_{\phi}^{(0)}+\xi H_{\phi}^{(2)}+\kappa H_{\phi}^{(4)},\hspace{1mm} H_{\mu}=H_{\mu}^{(2)}+\frac{\kappa}{\xi}H_{\mu}^{(4)},\hspace{1mm}
H_{\mu\nu}=H_{\mu\nu}^{(0)}+\xi H_{\mu\nu}^{(2)}+\kappa H_{\mu\nu}^{(4)}.
\end{align}
Here, the superscript (0) denotes the contribution due to JT gravity. On the other hand, the superscripts (2) and (4) denote the contributions due to 2-derivative and 4-derivative interaction terms.

The action constructed in equation (\ref{action 2D}) exhibits both vacuum solution as well as black hole solution. In this section, we study vacuum solution in detail. The general plan is to solve equations (\ref{ee}) at different order in perturbation as discussed in the following subsections.
\subsection{Zeroth order solutions}
In order to find out the vacuum solutions $\omega_{(0)}^{vac}$ and $\phi_{(0)}^{vac}$ , we set $\xi=\kappa=0$ in $\sqrt{-g}H_{\mu\nu}$, $\sqrt{-g}H_{\phi}$ and $\sqrt{-g}H^{\mu}$. This yields the following set of equations
\begin{eqnarray}
\phi_{(0)}''-\omega_{(0)}'\phi_{(0)}'-6\phi_{(0)}e^{2\omega_{(0)}}=0,\label{v01}\\
 \omega_{(0)}'\phi_{(0)}'-6\phi_{(0)}e^{2\omega_{(0)}}=0,\label{v02}\\
 12- 2e^{-2\omega_{(0)}}\omega_{(0)}'' = 0.\label{v03}
\end{eqnarray}

On solving (\ref{v01}), (\ref{v02}) and (\ref{v03}) we get
\begin{eqnarray}\label{v0s}
e^{2\omega_{(0)}^{vac}} =\frac{1}{6z^2}, \hspace{3mm}\phi_{(0)}^{vac}=-\frac{C_1}{z}
\end{eqnarray}
where $C_is$ are the integration constants. Equation (\ref{v0s}) stands for the vacuum solutions of pure JT gravity.  
\subsection{First order solutions in \texorpdfstring{$\xi$}{xi} }
Next, we note down leading order solutions (due to 2-derivative terms) by equating the coefficient of $\xi$ in $\sqrt{-g}H_{\mu\nu}$, $\sqrt{-g}H_{\phi}$ and $\sqrt{-g}H^{\mu}$ to zero
\begin{align}
2\Big(\omega_{(0)}'\phi_{(1)}'+\omega_{(1)}'\phi_{(0)}'+2\omega_{(1)}\phi_{(0)}'\omega_{(0)}'\Big)-(\phi_{(1)}''+2\omega_{(1)}\phi_{(0)}'')=0,\label{v21}\\
12e^{2\omega_{(0)}}\omega_{(1)}-\omega_{(1)}''+\frac{1}{4}e^{-2\omega_{(0)}} ({A_{t(0)}}')^2=0,\label{v22}\\
\partial_z\Big[\phi_{(0)}e^{-2\omega_{(0)}}{A_{t(0)}}'\Big]=0,\label{v23}
\end{align}
where (\ref{v21}) is corresponding to $\sqrt{-g}(H_{tt}+H_{zz})$. On solving equation (\ref{v21})-(\ref{v23}) we find, 
\begin{eqnarray}
&&A_{t(0)}^{vac}=-\frac{C_3}{C_1}\log z+C_4,\label{vs21}\\
&&\omega_{(1)}^{vac}=C_5z^2+\frac{C_6}{z}+\frac{z^2C_3^2(-1+3\log z )}{6C_1^2},\label{vs22}\\
&&\phi_{(1)}^{vac}= \frac{C_3^2z(-4+3\log z)}{3C_1}+C_1\Big(2C_5z-\frac{C_6}{z^2}\Big)-\frac{C_7}{z} + C_8,\label{vs23}
\end{eqnarray}
 To summarise, (\ref{vs21})-(\ref{vs23}) are the first order corrections to the pure JT gravity solutions due to 2-derivative interactions present in (\ref{action 2D}).
\subsection{First order solutions in \texorpdfstring{$\kappa$ }{kappa}}
Next, we note down leading order contributions due to the presence of 4-derivative interactions in (\ref{action 2D}). This can be calculated by equating the coefficients of $\kappa$ in $\sqrt{-g}H_{\mu\nu}$, $\sqrt{-g}H_{\phi}$ and $\sqrt{-g}H^{\mu}$ to zero.
\begin{align} 
&\partial_z\Big [-e^{-2\omega_{(0)}}\phi_{(0)} A_{t(1)}'-24e^{-6\omega_{(0)}}({A_{t(0)}}')^3\phi_{(0)}+ 2\partial_z\big\{e^{-4\omega_{(0)}}\phi_{(0)}(2\omega_{(0)}'{A_{t(0)}}'-{A_{t(0)}}'')\big\}\Big ]=0 ,\label{v41}\\
&24e^{2\omega_{(0)}}\omega_{(2)}-2\omega_{(2)}''+ \frac{24}{z^2}-\frac{5552}{27z^2}\Big(-\frac{C_1}{z}\Big)^{\frac{4}{3}}-\frac{40}{3z^2}\Big(-\frac{C_1}{z}\Big)^{\frac{8}{3}}
-\frac{3888z^2C_3^4}{C_1^4}-\nonumber\\
&\frac{C_3}{C_1^2}\Big(108C_3+C_9\Big)=0,\label{v42} \\
&6e^{2\omega_{(0)}}(\phi_{(2)}+4\omega_{(2)}\phi_{(0)})+(\phi_{(0)}'\omega_{(2)}'+\phi_{(2)}'\omega_{(0)}'+2\omega_{(2)}\phi_{(0)}'\omega_{(0)}')-(\phi_{(2)}''+2\omega_{(2)}\phi_{(0)}'')-\nonumber\\
&\frac{1}{18z^5C_1^3}\Bigg [216z^2C_1^4+976zC_1^5\Big(-\frac{C_1}{z}\Big)^{\frac{1}{3}}+8C_1^6\Big(-\frac{C_1}{z}\Big)^{\frac{2}{3}} +11664z^6C_3^4+
9z^4C_1^2C_3C_9\Bigg ]=0,\label{v43}\\
 &-6e^{2\omega_{(0)}}(\phi_{(2)}+4\omega_{(2)}\phi_{(0)})+(\phi_{(0)}'\omega_{(2)}'+\phi_{(2)}'\omega_{(0)}'+2\omega_{(2)}\phi_{(0)}'\omega_{(0)}')+\frac{1}{18z^3}\Bigg [216C_1+\nonumber\\
&24z\Big(-\frac{C_1}{z}\Big)^{\frac{11}{3}}+16z\Big(-\frac{C_1}{z}\Big)^{\frac{7}{3}}\Big\{-133+30\Big(-\frac{C_1}{z}\Big)^{\frac{1}{3}}\Big\}+11664z^4\frac{C_3^4}{C_1^3}+\nonumber\\
&9z^2\frac{C_3}{C_1}(72C_3+C_9)\Bigg]=0.\label{v44}
\end{align}

Notice that, (\ref{v44}) contains only single derivative terms which means that it is a constraint equation. We will use this constraint and equation (\ref{v0s}) in order to find $\phi_{(2)}^{vac}$ from equation (\ref{v43}). On the other hand, $A_{t(1)}^{vac}$ and $\omega_{(2)}^{vac}$ can be calculated using equations (\ref{v41}) and (\ref{v42}) respectively
\begin{align}
A_{t(1)}^{vac} = & \frac{432 z^2 C_3^3}{C_1^3}+\frac{\log z(72C_3+C_9)}{6C_1}+C_{10},\label{v4s1}\\
\omega_{(2)}^{vac} = & C_{11}z^2+\frac{C_{12}}{z}-6-\frac{1388}{15}\Big(-\frac{C_1}{z}\Big)^{\frac{4}{3}}-\frac{z^2}{18C_1^2}(-1+3\log z)C_3(108C_3+C_9)-\nonumber\\
&\frac{6}{7}\Big(-\frac{C_1}{z}\Big)^{\frac{8}{3}}-\frac{972}{5C_1^4}z^4C_3^4,\label{v4s2}\\
\phi_{(2)}^{vac} =& \frac{z^3C_{13}+C_{14}}{z}+\frac{2C_1^3}{7z^3}\Big(-\frac{C_1}{z}\Big)^{\frac{2}{3}}-\frac{12}{455}\Big(-\frac{C_1}{z}\Big)^{\frac{7}{3}}\Big\{2009+130\Big(-\frac{C_1}{z}\Big)^{\frac{1}{3}}\Big\}-\frac{648z^3C_3^4}{5C_1^3}\nonumber\\
&+C_1\Big(2zC_{11}-\frac{C_{12}}{z^2}\Big)+\frac{1}{9C_1}\Big[zC_3\{108(1-3\log z)C_3+
(4-3\log (z)C_9)\}\Big].\label{v4s3}
\end{align}
Equations (\ref{v4s1})-(\ref{v4s3}) are the first order corrections to pure JT gravity due to 4-derivative interactions in (\ref{action 2D}).

Now, we have a complete set of solutions corresponding to metric, gauge fields and dilaton up to linear order in $\xi$ and $\kappa$. Collecting all these fields at different order, we can approximate the space-time metric (\ref{sg}) for vacuum solution as
\begin{align}\label{VM}
    ds_{vac}^2\approx e^{2\omega_{(0)}^{vac}}(1+2\xi\omega_{(1)}^{vac}+2\kappa\omega_{(2)}^{vac})(-dt^2+dz^2).
\end{align}

Below, we check the behaviour of space-time metric in two different limits -
\begin{itemize}
    \item  Case 1 : IR limit i.e. $z\rightarrow\infty$
\begin{align}
e^{2\omega^{vac}}=&\frac{1}{6z^2}+\xi\Big\{\frac{-3C_3^2+9\log (z) C_3^2+18C_1^2C_5}{54C_1^2}+\frac{C_6}{3z^3}\Big\}+\kappa\Big\{-\frac{2}{z^2}-\frac{324z^2C_3^4}{5C_1^4}+\nonumber\\
&\frac{1}{54C_1^2}\Big(108C_3^2-324\log (z)C_3^2+C_3C_9-3\log(z)C_3C_9+18C_1^2C_{11}\Big)+\nonumber\\
&\frac{1}{45z^3}\Big(1388C_1\Big(-\frac{C_1}{z}\Big)^{\frac{1}{3}}+15C_{12}\Big)\Big\},\label{vir}
\end{align}
\item Case 2 : UV limit i.e. $z\rightarrow0$
\begin{align}
e^{2\omega^{vac}}=&\frac{1}{6z^2}+\xi\Big\{\frac{-3C_3^2+9\log(z)C_3^2+18C_1^2C_5}{54C_1^2}+\frac{C_6}{3z^3}\Big\}+\kappa\Big\{-\frac{2}{z^2}-\frac{2C_1^2(-C_1)^{\frac{2}{3}}}{7z^{\frac{14}{3}}}-\nonumber\\
&\frac{324z^2C_3^4}{5C_1^4}+\frac{1}{54C_1^2}\Big(108C_3^2-324\log(z)C_3^2+C_3C_9-3\log(z)C_3C_9+\nonumber\\
&18C_1^2C_{11}\Big)+\frac{1}{45z^3}\Big(1388C_1\Big(-\frac{C_1}{z}\Big)^{\frac{1}{3}}+15C_{12}\Big)\Big\}.\label{vuv}
\end{align}
\end{itemize}

It is evident from (\ref{vir}) and (\ref{vuv}) that the 2-derivative and 4-derivative interaction terms present in our model alter the $AdS_2$ geometry of vacuum both in the UV and IR limits. In the UV limit (\ref{vuv}), the space-time geometry is dominated by the Lifshitz$_2$ with dynamical exponent $z=\frac{7}{3}$. On the other hand, the space-time metric exhibits a divergence as we move in the deep IR limit (\ref{vir}). 

In order to solidify our claim, we further compute the corresponding scalar curvature of the theory (\ref{action 2D}) which shows a divergence in the deep IR namely,
$$R\big|_{z\rightarrow \infty}\sim\kappa\Big(\frac{C_3}{C_1}\Big)^4z^4,$$
where $\frac{C_3}{C_1}$ is precisely the coefficient that appears in the near boundary expansion of (\ref{VM}). This clearly reveals the fact that the space-time singularity is caused due to the presence of 4-derivative interactions in the original action (\ref{action 2D}). We identify this as the unique feature of higher derivative corrections in the theory (\ref{action 2D}).
\section{Boundary terms and renormalised action }\label{S4}
The boundary of space-time manifold in our theory (\ref{action 2D}) is located at $z=0$. Therefore, one must add  suitable boundary terms in action for a successful execution of variational principle \cite{Castro:2008ms}. 

The boundary term is given by standard Gibbons-Hawking-York term
\begin{eqnarray}\label{GHY}
S_{GHY}&=&D_1\int_0^{\beta} dt\sqrt{-\gamma}\phi K , \hspace{2mm} K=n^z\frac{\partial_z\sqrt{-\gamma}}{\sqrt{-\gamma}},\hspace{3mm}n^z=-\frac{1}{\sqrt{g_{zz}}},
\end{eqnarray} 
where $\gamma$ is the determinant of induced metric on boundary, $K$ is the trace of extrinsic curvature and $\beta$ is the inverse temperature \cite{Gibbons:1976ue}. We multiply the boundary term (\ref{GHY}) with an overall constant $D_1$ which will prove to be useful in construction of counter terms.

On substituting equation (\ref{VM}) in (\ref{GHY}), we obtain  $S_{GHY}=-\beta(\phi \omega')$. Using this expression, one can easily write down the on-shell Gibbons-Hawking-York boundary term as well as the on-shell Einstein-Hilbert action (\ref{action 2D}) as follows
\begin{align}
S_{GHY}^{on}=&-D_1 \beta \Bigg [\frac{C_1}{z^2}+\frac{103172\kappa  C_1^2}{585z^3}\Big(-\frac{C_1}{z}\Big)^{\frac{1}{3}}+\frac{24\kappa C_1^2}{7z^3}\Big(-\frac{C_1}{z}\Big)^{\frac{2}{3}}-\frac{18\kappa C_1^3}{7z^4}\Big(-\frac{C_1}{z}\Big)^{\frac{2}{3}}-\nonumber\\
&\frac{6\kappa C_3^2}{C_1}+\frac{7\xi C_3^2}{6C_1}+\frac{72\kappa \log(z)C_3^2}{C_1}-\frac{2\xi \log(z)C_3^2}{C_1}+\frac{4536z^2\kappa C_3^4}{5C_1^3}-4\xi C_1C_{5}+\nonumber\\
&\frac{2\xi C_1C_{6}}{z^3}+\frac{\xi C_7}{z^2}-\frac{\xi C_8}{z}-\frac{7\kappa C_3C_9}{18C_1}+\frac{2\kappa \log(z)C_3C_9}{3C_1}-4\kappa C_1 C_{11}+\frac{2\kappa C_1 C_{12}}{z^3}\nonumber \label{SGHY}\\
&-z\kappa C_{13}-\frac{\kappa C_{14}}{z^2}\Bigg],\\
S_{EH}^{on}=& -\frac{\beta \kappa}{27 C_1^3}\Bigg [\frac{6096C_1^5}{5z^3}\Big(-\frac{C_1}{z}\Big)^{\frac{1}{3}}+\frac{576C_1^5}{7z^3}\Big(-\frac{C_1}{z}\Big)^{\frac{5}{3}}+69984z^2C_3^4 +27\log(z)C_1^2C_3\times\nonumber\\
&(72C_3+C_9) \Bigg ],\label{SEH}
\end{align}
where we have truncated the above expressions (\ref{SGHY}) and (\ref{SEH}) up to linear order in $\xi$ and $\kappa$.

It should be noted that in the boundary limit i.e. $z\rightarrow 0$, both the equations (\ref{SGHY}) and (\ref{SEH}) diverge. Therefore, one requires to add counter terms in the action (\ref{action 2D}) to tame such UV divergences. These counter terms should be some function of the fields at boundary.

After a careful inspection, we come up with the following counter term
\begin{eqnarray}\label{SCT}
S_{CT}&=& \int_{0}^{\beta}dt\sqrt{-\gamma}\Bigg[D_2\phi+C_1D_3\sqrt{-\gamma}K^2+C_3D_4\xi\sqrt{-\gamma^{\mu\nu}A_{\mu}A_{\nu}}+\xi \frac{D_5}{\sqrt{-\gamma}}\frac{C_6}{C_1^2}\phi^3 \nonumber\\
&&+\kappa \frac{D_6}{\sqrt{-\gamma}}\frac{C_{12}}{C_1^2}\phi^3\Bigg ],
\end{eqnarray}
where $C_is$ and $D_is$ are some constant coefficients.

Equation (\ref{SCT}) cures all the UV divergences of $S_{EH}^{on}+S_{GHY}^{on}$ (up to linear order in $\xi$ and $\kappa$) with a particular choice of coefficients\footnote{See Appendix \ref{const} for a detailed derivation of the coefficients. }
\begin{align}
    &D_1=-2.5795,\hspace{1mm}D_2=6.3186,\hspace{1mm}D_3=-0.1394,\hspace{1mm}D_4=-3.5904,\hspace{1mm}D_5=D_6=-0.2788,\hspace{1mm}\nonumber\\ &C_3=0.0283C_9,\hspace{1mm}\kappa=0.0090.
\end{align}

Notice that, in the process of renormalization we also fixed the value of 4-derivative coupling constant, $\kappa$. With all these preliminaries, the complete renormalised  action can be schematically expressed as
\begin{align}\label{S2D}
    S_{2D}=S_{EH}+S_{GHY}+S_{CT}.
\end{align}

The variation of the full action (\ref{S2D}) is given by
\begin{align}\label{VS2D}
\delta S_{2D}=\int dt\sqrt{-\gamma}\Big[G^{ab}\delta \gamma_{ab}+G_{\phi}\delta{\phi}+G^a\delta A{_a}\Big ] +\hspace{1mm} bulk \hspace{2mm}terms
\end{align} 
where  (a,b) are the boundary indices. 

The bulk part of (\ref{VS2D}) is already discussed in Sections (\ref{S2}) and (\ref{S3}). On the other hand, the variation of boundary action yields
\begin{eqnarray}
G_{\phi}&=&D_1K+D_2+3\xi\frac{D_5}{\sqrt{-\gamma}}\frac{C_6}{C_1^2}\phi^2+3\kappa\frac{D_6}{\sqrt{-\gamma}}\frac{C_{12}}{C_1^2}\phi^2 ,\\
G^t&=&-\xi n_{\alpha} F^{\alpha t}\phi-\xi C_3 D_4\frac{\gamma^{tt}A_t}{\sqrt{-\gamma^{tt}A_tA_t}},\\
G^{tt}&=&\frac{1}{\sqrt{g_{zz}}}\Big\{\partial_z\gamma^{tt}\phi-\partial_z\phi\gamma^{tt}+\frac{2}{z}\gamma^{tt}\phi\Big\}+\frac{D_1}{2}\frac{\phi}{\sqrt{g_{zz}}}\Big\{-\partial_z\gamma^{tt}-\frac{2}{z}\gamma^{tt}-\frac{1}{\sqrt{-\gamma}}\gamma^{tt}\partial_z\sqrt{-\gamma}\Big\}\nonumber\\
&&+\frac{D_2}{2}\gamma^{tt}\phi+C_1D_3\frac{K}{\sqrt{g_{zz}}}\Big\{-\partial_z\gamma^{tt}\sqrt{-\gamma}-\frac{2}{z}\gamma^{tt}\sqrt{-\gamma}-\gamma^{tt}\partial_z\sqrt{-\gamma}\Big\}+\nonumber\\
&& C_3D_4\frac{\xi}{2}\Bigg\{\gamma^{tt}\sqrt{-\gamma^{tt}A_t A_t}-\frac{A^t A^t}{\sqrt{-\gamma^{tt}A_tA_t}}\Bigg\}.\label{bgtt}
\end{eqnarray}

In arriving at equation (\ref{bgtt}), we have used (\ref{GHY}) and the dominating terms in the expansion of $\delta\gamma_{tt}$ near boundary. It is important to note that the above variation (\ref{VS2D}) makes sense only when the individual variations of the metric ($\delta\gamma_{ab}$), dilaton $(\delta\phi)$ and the gauge ($\delta A_a$) field vanishes at the boundary.

In order to check this explicitly we expand the variation of all fields near boundary which yield
\begin{eqnarray}
\delta\phi&=&\frac{1}{9C_1}\Big[z\{3(36\kappa-4\xi+3(-36\kappa+\xi)\log(z))C_3^2+\kappa(4-3\log(z))C_3C_9+\nonumber \\
&&18C_1^2(\xi C_5+\kappa C_{11})\}\Big ]+O[z]^2,\label{ve1}\\
\delta A_t&=&432\frac{C_3^3}{C_1^3}\frac{\kappa}{\xi}z^2+O[z]^3\hspace{2mm},\hspace{5mm}\delta\gamma_{tt}= \kappa \frac{324}{5}\frac{C_3^4}{C_1^4}z^2+O[z]^3.\label{ve2}
\end{eqnarray}

From equations (\ref{ve1}) and (\ref{ve2}), it is quite evident that the individual variation of fields $\delta\phi$, $\delta A_t$ and $\delta \gamma_{tt}$ vanishes in the boundary limit $z\rightarrow 0$.  Therefore, the results derived above are all reliable and we will use them in deriving the boundary stress tensor in the next section.
\section{Stress tensor and central charge}\label{S5}
Having done the required background work, we now proceed towards computing the stress  tensor as well as the central charge for the boundary theory. Boundary stress tensor is defined as the variation of the action (\ref{S2D}) with respect to the induced metric ($\gamma_{ab}$)
\begin{eqnarray}\label{std}
T^{ab}&=&\frac{2}{\sqrt{-\gamma}}\frac{\delta S_{2D}}{\delta\gamma_{ab}}=2G^{ab},
\end{eqnarray}
where $G^{ab}$ is given by equation (\ref{bgtt}).

So far, our computations have been performed in the light cone gauge (\ref{sg}). However, it is not convenient to identify the central charge in this gauge. Therefore, we switch to so called Fefferman Graham gauge \cite{fefferman} in which it is quite straightforward to figure out the central charge.
\subsection{The Fefferman-Graham gauge }
In this section, we will demonstrate how to write down the background fields in the Fefferman-Graham gauge. In order to do that, we first make a coordinate transformation that takes us into the Fefferman-Graham gauge from the light cone gauge. This can be done as follows.

Consider the line element in light cone gauge   
\begin{eqnarray}\label{lm}
ds^2=-e^{2\omega(z)}dt^2+e^{2\omega(z)}dz^2.
\end{eqnarray}

Now, consider the following transformation
\begin{eqnarray}
d\eta &=&e^{\omega(z)}dz,
\end{eqnarray}
which by virtue of (\ref{e1}) and (\ref{v0s}) yields
\begin{eqnarray}\label{fgge}
\eta &=&\int\frac{1}{\sqrt{6}z}(1+\xi \omega_{(1)}+\kappa\omega_{(2)})dz.
\end{eqnarray}

 In principle, one can evaluate (\ref{fgge}) using equation (\ref{vs22}) and (\ref{v4s2}).  This will give us $\eta$ as a function of $z$ i.e. $\eta\equiv\eta(z)$. One can therefore revert (\ref{fgge}) to express $z$ as a function of $\eta$  and plug it back into equation (\ref{lm}). This yields the desired form of the line element in the Fefferman-Graham gauge
\begin{eqnarray}\label{fgm}
ds^2=h_{tt}(\eta)dt^2+d\eta^2.
\end{eqnarray}

In order to simplify our analysis further, we expand equation (\ref{fgge}) in the boundary limit ($z\rightarrow0$) and retain only dominating terms in the expansion. Notice that, the boundary in the  Fefferman-Graham gauge is located at $\eta=\infty$. 

Upon solving equation (\ref{fgge}) and expressing $z$ as a function of $\eta$ we get 
\begin{eqnarray}\label{zfg}
z= \frac{3^{\frac{9}{16}}\kappa^{\frac{3}{8}}\tilde{C_1}}{2^{\frac{15}{16}}\times7^{\frac{3}{8}}\eta^{\frac{3}{8}}},
\end{eqnarray}
where $\tilde{C_1}=-C_1$. The above expression (\ref{zfg}) will be used while converting the light cone gauge into the Fefferman-Graham gauge and vice-versa. 

Using  (\ref{lm}), (\ref{zfg}) and (\ref{e1})-(\ref{e3}) we finally end up with the following expressions for the background fields as well as the stress tensor in the Fefferman-Graham gauge
\begin{eqnarray}
h_{tt}(\eta)\Big|_{\eta\rightarrow\infty}&=&\frac{32\times2^{\frac{3}{8}}\times 7^{\frac{3}{4}}\times\eta^{\frac{3}{2}}\Big(\frac{\eta^{\frac{3}{8}}}{\kappa^{\frac{3}{8}}}\Big)^{\frac{2}{3}}}{9\times3^{\frac{5}{8}}\sqrt{\kappa}\tilde{C_1}^2}+...,\label{fgh}\\
\phi(\eta)\Big|_{\eta\rightarrow\infty}&=&-\frac{16\times2^{\frac{7}{16}}\times 7^{\frac{3}{8}}\times\eta^{\frac{9}{8}}\Big(\frac{\eta^{\frac{3}{8}}}{\kappa^{\frac{3}{8}}}\Big)^{\frac{2}{3}}}{9\times3^{\frac{1}{16}}\kappa^{\frac{1}{8}}}+...,\label{fgp}\\
A_t(\eta)\Big|_{\eta\rightarrow\infty}&=&-\log\Bigg(\frac{3^{\frac{9}{16}}\kappa^{\frac{3}{8}}\tilde{C_1}}{2^{\frac{15}{16}}\times7^{\frac{3}{8}}\eta^{\frac{3}{8}}}\Bigg)\Bigg(\frac{((72\kappa-6\xi)C_3+\kappa C_9)}{6\xi\tilde{C_1}}\Bigg)+..,\label{fga}\\
T_{tt}(\eta)\Big|_{\eta\rightarrow\infty}&=& -3175.934\frac{\eta^{3.5}}{\kappa^{1.5}\tilde{C_1}^3}-294.245\frac{\eta^{3.125}}{\kappa^{1.125}\tilde{C_1}^2}+...\label{fgt}
\end{eqnarray}
where (...) represents all the sub leading terms in an expansion near the boundary.
\subsection{Transformation properties of the stress tensor}
In the present Section, we study the transformation properties of the boundary stress tensor under diffeomorphism. Under diffeomorphism, $x^{\mu}\rightarrow x^{\mu}+\epsilon^{\mu}(x)$ the space time metric, gauge fields and the dilaton transform as follows
\begin{eqnarray}
\delta_{\epsilon}g_{\mu\nu}&=&\bigtriangledown_{\mu}\epsilon_{\nu}+\bigtriangledown_{\nu}\epsilon_{\mu},\label{df1}\\
\delta_{\epsilon}A_{\mu}&=&\epsilon^{\nu}\bigtriangledown_{\nu}A_{\mu}+A_{\nu}\bigtriangledown_{\mu}\epsilon^{\nu},\label{df2}\\
\delta_{\epsilon}\phi&=&\epsilon^{\mu}\bigtriangledown_{\mu} \phi.\label{df3}
\end{eqnarray}

Using (\ref{fgm}), (\ref{fgh}) and (\ref{df1}), one can find an expression for the parameter ($\epsilon_{\mu}$) of diffeomorphism which turns out to be
\begin{eqnarray}\label{dfp}
\epsilon_{\eta}= a\Xi'(t),\hspace{2mm}\epsilon_t=b\eta^{\frac{7}{4}}\Xi(t)+\frac{4}{3}a\eta\Xi''(t),
\end{eqnarray}
where $\Xi(t)$ is an arbitrary function of time while the constants (a, b) will be fixed latter on.

Recall, that we are working in a gauge in which $A_{\eta}$ is set to be zero. From equations (\ref{df2}) and (\ref{dfp}), it is easy to check that $\delta_{\epsilon}A_{\eta}\neq 0$ which means that the differomorphism destroys the gauge condition. Therefore, to retain the gauge condition, we make another gauge transformation i.e.  $A_{\mu}\rightarrow A_{\mu}+\partial_{\mu}\lambda$, where we choose $\lambda$ such that $(\delta_{\epsilon}+\delta_{\lambda})A_{\eta}=0$, which determines $\lambda$ at leading order as
\begin{eqnarray}\label{flam}
\lambda&=&-\frac{1}{256\times2^{\frac{3}{8}}\times7^{\frac{3}{4}}\eta^{\frac{7}{8}}\xi}\Bigg[3^{\frac{5}{8}}a\Big(\frac{\eta^{\frac{3}{8}}}{\kappa^{\frac{3}{8}}}\Big)^{\frac{1}{3}}\kappa^{\frac{7}{8}}\Big(8+15\log(2)-9\log(3)+6\log(7)\nonumber\\
&&-16\log\Big(\frac{\kappa^{\frac{3}{8}} \tilde{C_1}}{\eta^{\frac{3}{8}}}\Big)\Big)\tilde{C_1}((72\kappa-6\xi)C_3+\kappa C_9)\Xi''(t)\Bigg].
\end{eqnarray}

Using (\ref{dfp}) and (\ref{flam}), one can finally pin down the variations of the background fields under the diffeomorphism and the gauge transformation as
\begin{align}
\delta_{\epsilon}h_{tt}&=\frac{2}{27}\sqrt{\eta}\Big(27b\eta^{\frac{5}{4}}+\frac{28\times6^{\frac{3}{8}}\times7^{\frac{3}{4}}a}{\sqrt{\kappa}\tilde{C_1}^2}\Big(\frac{\eta^{\frac{3}{8}}}{\kappa^{\frac{3}{8}}}\Big)^{\frac{2}{3}}\Big)\Xi'(t)+\frac{8}{3}a\eta\Xi'''(t),\\
(\delta_{\epsilon}+\delta_{\lambda})A_t&=\frac{1}{896\eta\Big(\frac{\eta^{\frac{3}{8}}}{\kappa^{\frac{3}{8}}}\Big)^{\frac{2}{3}}\xi\tilde{C_1}}((72\kappa-6\xi)C_3+\kappa C_9)\Bigg[\Big(56a\Big(\frac{\eta^{\frac{3}{8}}}{\kappa^{\frac{3}{8}}}\Big)^{\frac{2}{3}}-3\times6^{\frac{5}{8}}\times7^{\frac{1}{4}}b\eta^{\frac{5}{4}}\sqrt{\kappa}\nonumber\\
&\times \log\Big(\frac{3^{\frac{9}{16}}\kappa^{\frac{3}{8}}\tilde{C_1}}{2^{\frac{15}{16}}\times7^{\frac{3}{8}}\eta^{\frac{3}{8}}}\Big)\tilde{C_1}^2\Big)\Xi'(t)-2\times6^{\frac{5}{8}}\times7^{\frac{1}{4}}a\sqrt{\eta}\sqrt{\kappa}\tilde{C_1}^2\Xi'''(t)\Bigg],\\
\delta_{\epsilon}\phi&=-\frac{22\times2^{\frac{7}{16}}\times7^{\frac{3}{8}}a\eta^{\frac{1}{8}}\Xi'(t)}{9\times3^{\frac{1}{16}}\kappa^{\frac{1}{8}}}\Big(\frac{\eta^{\frac{3}{8}}}{\kappa^{\frac{3}{8}}}\Big)^{\frac{2}{3}}.
\end{align}
 
 In order to proceed further, we first convert the stress tensor (see equation (\ref{std}) and (\ref{bgtt})) into  Fefferman-Graham coordinate and then explore its properties under diffeomorphism and gauge transformation. After doing all the calculations, we end up with the following expression
\begin{align}\label{dst1}
(\delta_{\epsilon}+\delta_{\lambda})T_{tt}&\approx \Big(-58.928 b\frac{\eta^{3}}{\kappa^{0.375}}-636.04b\frac{\eta^{3}}{\kappa^{0.75}\tilde{C_1}}\Big)\Xi'(t)
-672.835a\frac{\eta^{3}}{\kappa^{0.75}\tilde{C_1}}\Xi'''(t),
\end{align}
where we have retained only the dominant terms in the (boundary) limit $\eta\rightarrow \infty$. 

After a proper re-scaling, the boundary stress tensor and its variation under diffeomorphism and gauge transformation may be defined as,
\begin{eqnarray}\label{rst}
\tilde{T_{tt}}=\lim_{\eta\rightarrow\infty}\frac{1}{\eta^3}T_{tt}\hspace{2mm}and\hspace{2mm}(\delta_{\epsilon}+\delta_{\lambda})\tilde{T_{tt}}=\lim_{\eta\rightarrow\infty}\frac{1}{\eta^3}(\delta_{\epsilon}+\delta_{\lambda})T_{tt}.
\end{eqnarray}

With a proper choice of the constant $b =\frac{9.986}{\tilde{C_1}^2\kappa^{0.75}}$, one can express the equation (\ref{rst}) in a more elegant way. Using (\ref{fgt}), this finally leads to the transformation of the stress tensor\footnote{Since we are working in a static gauge therefore, $\Xi(t)\partial_t\tilde{T_{tt}}$ is trivially zero. } as follows
\begin{eqnarray}\label{cdm1}
(\delta_{\epsilon}+\delta_{\lambda})\tilde{T_{tt}}=2\tilde{T_{tt}}\Xi'(t)- \frac{ca}{\tilde{C_1}}\Xi'''(t).
\end{eqnarray}
  
  (\ref{cdm1}) is the standard form of variation of the boundary stress tensor (\ref{std}) under the action of  both diffeomorphism and gauge transformation. Finally, we have reached a stage where one can identify the central charge of the boundary theory. The constant $``c"$ appearing in (\ref{cdm1}) (as the coefficient of $\Xi'''(t)$) is the central charge associated to our boundary theory (\ref{S2D}) which is given by the following expression
 \begin{align}\label{cc}
     c=\frac{672.835}{\kappa^{0.75}}.
 \end{align}

Notice that, (\ref{cc}) is a large number as we are working in the small $\kappa$ regime. This also makes the entity in (\ref{cc}) highly non-perturbative in the sense that there does not exist any smooth  $\kappa\rightarrow 0 $ limit of (\ref{cc}) that connects it to the pure JT gravity theory. Therefore, these theories are not smoothly connected to their conformal cousins those are dual to pure JT gravity.
\section{Black hole solutions}\label{SBH}
We now explore black hole solutions of the 2D gravity model (\ref{action 2D}). Like before, these solutions are expressed perturbatively with the gauge choice as discussed in Section (\ref{S3}).
\subsection{Zeroth order solution }
In order to calculate the zeroth order solution, we solve equations (\ref{v01}), (\ref{v02}) and (\ref{v03}) simultaneously which yields
\begin{eqnarray}\label{bh0}
e^{2\omega_{(0)}^{bh}}=\frac{8\mu}{12[\sinh(2\sqrt{\mu} z)]^2}\hspace{1mm},\hspace{2mm}\phi_{(0)}^{bh}=\frac{\sqrt{\mu}}{6}\coth(2z\sqrt{\mu}).
\end{eqnarray}
The above solutions (\ref{bh0}) correspond to black hole solutions in pure JT gravity \cite{Almheiri:2014cka}.
\subsection{First order corrections in \texorpdfstring{$\xi$}{xi}}
Leading order corrections to (\ref{bh0}) can be estimated by using equations (\ref{v21})-(\ref{v23}). These equations will be easy to handle if we change the coordinate as follows
\begin{eqnarray}
z=\frac{1}{2\sqrt{\mu}}\coth^{-1}\Big(\frac{\rho}{\sqrt{\mu}}\Big).
\end{eqnarray}\label{ze}

Using (\ref{bh0}) and (\ref{ze}), we can express the first order solution as 
\begin{align}
A_{t(0)}^{bh}=&-2Q\log(\rho)+d_1,\label{bh21}\\
\omega_{(1)}^{bh}=&\frac{1}{4\mu^{\frac{3}{2}}}\Bigg[3\Big\{-2\rho \tanh^{-1}\Big(\frac{\rho}{\sqrt{\mu}}\Big)\log(\rho)+2\sqrt{\mu}(1+\log(\rho))-\rho \text{PolyLog}\Big[2,-\frac{\rho}{\sqrt{\mu}}\Big]\nonumber \\ 
&+\rho \text{PolyLog}\Big[2,\frac{\rho}{\sqrt{\mu}}\Big]\Big\}Q^2+4\mu \Big\{\rho d_1-\sqrt{\mu}d_2+\rho \tanh^{-1}\Big(\frac{\rho}{\sqrt{\mu}}\Big)d_2\Big\}\Bigg], \label{bh22}
\end{align}
\begin{align}
\phi_{(1)}^{bh}=&\frac{1}{48\mu^{\frac{3}{2}}}\Bigg[3\Big\{2\mu+4\sqrt{\mu}\rho+4\sqrt{\mu}\rho \log(\rho)+\tanh^{-1}\Big(\frac{\rho}{\sqrt{\mu}}\Big)\Big\{6\rho^2-8\sqrt{\mu}\rho +\nonumber\\
&4(\mu-\rho^2)\log(\rho)\Big\} -\mu \log\Big(1-\frac{\rho}{\sqrt{\mu}}\Big)-8\sqrt{\mu}\rho \log\Big(1-\frac{\rho}{\sqrt{\mu}}\Big)+6\rho^2\log\Big(1-\frac{\rho}{\sqrt{\mu}}\Big)\nonumber\\
&-\mu \log\Big(1+\frac{\rho}{\sqrt{\mu}}\Big)+4\sqrt{\mu}\rho \log(-\mu+\rho^2)+(\mu-3\rho^2)\log\Big(1-\frac{\rho^2}{\mu}\Big)-\nonumber \\
&(\mu-\rho^2)\Big(4\text{PolyLog}\Big(2,\frac{\rho}{\sqrt{\mu}}\Big)-\text{PolyLog}\Big(2,\frac{\rho^2}{\mu}\Big)\Big)\Big\}Q^2+8\mu \rho^2d_1+4\mu\Big\{2\sqrt{\mu}\rho+\nonumber \\
&2\rho^2\tanh^{-1}\Big(\frac{\rho}{\sqrt{\mu}}\Big)+\mu \log(-\sqrt{\mu}+\rho)-\mu \log(\sqrt{\mu}+\rho)\Big\}d_2+48\mu^{\frac{3}{2}}(d_3+\rho d_4)\Bigg]\label{bh23},
\end{align}
where $Q$ is the charge of the $U(1)$ gauge theory and $d_is$ are the constants where i takes the value 1, 2, 3, ... Equations (\ref{bh21}), (\ref{bh22}) and (\ref{bh23}) correspond to first order corrections to zeroth order (black hole) solutions due to 2-derivative interaction terms in (\ref{action 2D}).
\subsection{First order corrections in \texorpdfstring{$\kappa$}{kappa}}
Let us first estimate corrections to gauge fields due to 4-derivative interactions in (\ref{action 2D}). These can be estimated by comparing the coefficient of $\kappa$ in equation of motion for $A_{\mu}$ 
\begin{eqnarray}\label{bh41}
-e^{-2\omega_{(0)}}\phi_{(0)}A_{t(1)}'-24e^{-6\omega_{(0)}}\phi_{(0)}(A_{t(0)}')^3+2\partial_z[e^{-4\omega_{(0)}}\phi_{(0)}(2\omega_{(0)}'A_{t(0)}'-A_{t(0)}'')]=d_6.
\end{eqnarray}

Notice that the above equation (\ref{bh41}) is expressed in terms of $z$ and its derivatives. Upon solving equation (\ref{bh41}) in terms of $\rho$ and using equation (\ref{bh0}) we finally obtain
\begin{eqnarray}\label{bh4a}
A_{t(1)}^{bh}=\frac{12Q(\mu-72Q^2)}{\rho^2}+2\log(\rho)(12Q+d_5)+d_6.
\end{eqnarray}

Next, we collect the coefficient of $\kappa$ in equation of motion for $\phi$. After simplifying the expression we get,
\begin{align}\label{bh42}
&\frac{4}{243}(\mu-\rho^{2})\Bigg[-5832+\frac{6^{1 / 3}}{\rho^{\frac{8}{3}}}\Big\{-\mu^{2}(40 \times 6^{1 / 3}+\rho^{4 / 3})-2 \mu \rho^{2}(116 \times 6^{1 / 3}+7 \rho^{4 / 3})+\nonumber \\
&\rho^{4}(1388 \times 6^{1 / 3}+15 \rho^{4 / 3})\Big\}-972 \omega_{(2)}+\frac{486}{\rho^4}\Big\{54(-\mu+\rho^{2}) Q^{2}+1944 Q^{4}+3 \rho^{2} Q d_{5}+\nonumber\\
&\rho^{4}\Big(2 \rho \frac{\partial\omega_{(2)}}{\partial\rho}+(-\mu+\rho^{2}) \frac{\partial^2\omega_{(2)}}{\partial \rho^2}\Big)\Big\}\Bigg]=0.
\end{align}

In general, one can solve (\ref{bh42}) exactly for $\omega_{(2)}$. However, for the purpose of our present analysis, we are interested in the near boundary expression of this function. Therefore, we expand $\omega_{(2)}$ in the limit $\rho \rightarrow \infty$ and retain only leading order terms.

After simplification, one can express $\omega_{(2)}$ in the following form
\begin{align}
\omega_{(2)}^{bh}|_{\rho\rightarrow\infty}=-0.0072108\rho^{\frac{8}{3}}-8.48718\rho^{\frac{4}{3}} + F\big(\log(\mu),\log(\rho)\big)\rho- 0.0228342\mu\rho^{\frac{2}{3}}-6,  
\end{align}
where $F\big(\log(\mu),\log(\rho)\big)$ is given by
\begin{align}
    F=&\hspace{1mm}3.72651\mu^{\frac{1}{6}}-3.72651\Big(-\frac{1}{\sqrt{\mu}}\Big)^{\frac{2}{3}}\sqrt{\mu}+0.0239781\mu^{\frac{5}{6}}+0.0239781\Big(-\frac{1}{\sqrt{\mu}}\Big)^{\frac{1}{3}}\mu+\nonumber\\
    &\frac{3.375}{\mu^\frac{3}{2}}\Big\{4\log\Big(-\frac{1}{\sqrt{\mu}}\Big)^2-\log(\mu)^2+8\log\Big(-\frac{1}{\sqrt{\mu}}\Big)\log(\rho)+4\log(\mu)\log(\rho)-\nonumber\\
    &8\log\Big(-\frac{\rho}{\sqrt{\mu}}\Big)-8\log(\rho)\log\Big(-\frac{\rho}{\sqrt{\mu}}\Big)+8\log\Big(\frac{\rho}{\sqrt{\mu}}\Big)+8\log(\rho)\log\Big(\frac{\rho}{\sqrt{\mu}}\Big)\Big\}Q^2+\nonumber\\
    &\frac{972}{\mu^\frac{5}{2}}\Big(\log\Big(-\frac{\rho}{\sqrt{\mu}}\Big)Q^4-\log\Big(\frac{\rho}{\sqrt{\mu}}\Big)Q^4\Big)-\frac{0.1875Qd_5}{\mu^\frac{3}{2}}\Big\{-4\log\Big(-\frac{1}{\sqrt{\mu}}\Big)^2+\log(\mu)^2-\nonumber\\
    &8\log\Big(-\frac{1}{\sqrt{\mu}}\Big)\log(\rho)-4\log(\mu)\log(\rho)+8\log(\rho)\log\Big(-\frac{\rho}{\sqrt{\mu}}\Big)-8\log(\rho)\log\Big(\frac{\rho}{\sqrt{\mu}}\Big)\Big\}\nonumber\\
    &+\frac{d_7}{\sqrt{\mu}}+\frac{0.5}{\sqrt{\mu}}\Big(-\log\Big(\frac{-\rho}{\sqrt{\mu}}\Big)d_8+\log\Big(\frac{\rho}{\sqrt{\mu}}\Big)d_8\Big).
\end{align}

With all these expressions at hand, one can approximate the black hole metric (\ref{sg}) as 
\begin{eqnarray}\label{2dbh}
ds^2_{bh}=\frac{2}{3}(\rho^2-\mu)\Big(1+2(\xi \omega_{(1)}^{bh}+\kappa \omega_{(2)}^{bh})\Big)\Bigg(-dt^2+\frac{d\rho^2}{4(\mu-\rho^2)^2}\Bigg),
\end{eqnarray}
where the black hole horizon is located at $\rho=\sqrt{\mu}$.

In order to calculate $\phi_{(2)}$, we compare the coefficient of $\kappa$ in equations of motion of $g_{tt}$ and $g_{zz}$. On Subtracting $g_{zz}$ from $g_{tt}$ and after some simplification we find
\begin{align}\label{bh43}
12e^{2\omega_{(0)}}(\phi_{(2)}+4\phi_{(0)}\omega_{(2)})-\Big(\frac{\partial\rho}{\partial z}\frac{\partial}{\partial \rho}\Big\{\frac{\partial\rho}{\partial z}\Big(\frac{\partial \phi_{(2)}}{\partial \rho}\Big)\Big\}+2\omega_{(2)}\frac{\partial\rho}{\partial z}\frac{\partial}{\partial \rho}\Big\{\frac{\partial\rho}{\partial z}\Big(\frac{\partial \phi_{(0)}}{\partial \rho}\Big)\Big\}\Big)\nonumber\\+A(\rho)=0,
\end{align}
where $A(\rho)$ is given by 
\begin{align}
   A(\rho)&= \frac{2(\mu-\rho^2)}{729\rho^{\frac{1}{3}}}\Big\{6^{\frac{1}{3}}\mu^2-5832\rho^{\frac{4}{3}}-432\times6^{\frac{2}{3}}\rho^{\frac{8}{3}}+180\times6^{\frac{1}{3}}\rho^3+6^{\frac{1}{3}}\rho^4+2\mu(54\times6^{\frac{2}{3}}\rho^{\frac{2}{3}}-\nonumber\\
    &90\times6^{\frac{1}{3}}\rho-6^{\frac{1}{3}}\rho^2)\Big\}+\frac{4Q}{\rho^3}\Big\{6Q(4\mu^2-6\mu\rho^2+\rho^4+36(-\mu+\rho^2)Q^2)+\rho^2(-\mu+\rho^2)d_5\Big\}.
\end{align}

Technically speaking, it is very difficult to solve (\ref{bh43}) exactly. Therefore we will solve this equation in two different limits.

\begin{itemize}
    \item 
\underline{Case 1} : \textbf{Near boundary analysis} ($\rho\rightarrow\infty$) :\\
Using the near boundary expansion of $\omega_{(2)}$ and $A(\rho)$ in equation  (\ref{bh43}) we get
\begin{eqnarray}\label{BH44}
4\rho^2\frac{d^2\phi_{(2)}}{d\rho^2}+8\rho\frac{d\phi_{(2)}}{d\rho}-8\phi_{(2)}+0.02\rho^{\frac{11}{3}}=0,
\end{eqnarray}
where we consider $\rho^2>>\mu$ and retain only dominant terms in the above expression. Equation (\ref{BH44}) can be easily solved for $\phi_{(2)}$ 
\begin{eqnarray}\label{bh45}
\phi_{(2)}^{(bh)}|_{\rho\rightarrow\infty}=-\frac{9}{27200}\rho^{\frac{11}{3}}+\rho d_{9}+\frac{d_{10}}{\rho^2}.
\end{eqnarray}
\item \underline{Case 2} : \textbf{Near horizon analysis} ($\rho\rightarrow \sqrt{\mu}$) :\\
Converting equation (\ref{bh43}) in terms of $\rho$ and taking the limit $\rho\rightarrow\sqrt{\mu}$, we arrive at the following equation
\begin{eqnarray}
12e^{2\omega_0}\phi_2-\frac{\partial\rho}{\partial z}\frac{\partial}{\partial \rho}\Big\{\frac{\partial\rho}{\partial z}\Big(\frac{\partial \phi_2}{\partial \rho}\Big)\Big\}-24\sqrt{\mu}Q^2=0,
\end{eqnarray}
which can be solved for $\phi_{(2)}$ to yield,
\begin{align}\label{bh46}
\phi_{(2)}^{(bh)}|_{\rho\rightarrow\sqrt{\mu}}=&\frac{1}{\sqrt{\mu}}\Big(\rho d_{11}-\sqrt{\mu}d_{12}+\rho \tanh^{-1}\Big(\frac{\rho}{\sqrt{\mu}}\Big)d_{12}\Big)+\frac{3}{4\mu}\Big\{\rho\Big(4\log(-\sqrt{\mu}+\rho)-\nonumber\\
&4\log(\sqrt{\mu}+\rho)+\log\Big(1-\frac{\rho}{\sqrt{\mu}}\Big)^2+2\log\Big(1-\frac{\rho}{\sqrt{\mu}}\Big)\log\Big[\frac{1}{4}\Big(1+\frac{\rho}{\sqrt{\mu}}\Big)\Big]-\nonumber\\
&\log\Big(1+\frac{\rho}{\sqrt{\mu}}\Big)^2\Big)-4\Big\{\sqrt{\mu}-\rho \tanh^{-1}\Big(\frac{\rho}{\sqrt{\mu}}\Big)\Big\}\log(\mu-\rho^2)+\nonumber\\
&4\rho \text{PolyLog}\Big(2,\frac{1}{2}-\frac{\rho}{2\sqrt{\mu}}\Big)\Big\}Q^2.
\end{align}
\end{itemize}
The above set of solutions (\ref{bh4a})-(\ref{bh46}) are the first order corrections to pure JT gravity black hole solutions due to 4-derivative interaction terms in (\ref{action 2D}).

 Now, we have obtained a complete set of black hole as well as vacuum solutions for generalized JT gravity models with an abelian one form. Our next task would be to compare these solutions in the near boundary limit. Let us first expand the black hole solutions (\ref{bh0})-(\ref{bh46}) in the limit $z\rightarrow0$, which reveals the following leading order behaviour for the background fields and the metric 
 \begin{eqnarray}\label{bh47}
\phi^{(bh)}\Big|_{z\rightarrow 0}\sim\frac{1}{z^{\frac{11}{3}}}\hspace{1mm},\hspace{2mm}A_{t}^{(bh)}\Big|_{z\rightarrow 0}\sim \log(z)\hspace{1mm},\hspace{2mm}e^{2\omega ^{(bh)}}\Big|_{z\rightarrow 0}\sim \frac{1}{z^{\frac{14}{3}}}.
\end{eqnarray}

On the other hand, for vacuum solutions (\ref{v0s})-(\ref{v4s3}), we find the leading order behaviour for the background fields as well as the metric
\begin{eqnarray}\label{bh48}
\phi^{(vac)}\Big|_{z\rightarrow 0}\sim\frac{1}{z^{\frac{11}{3}}}\hspace{1mm},\hspace{2mm}A_{t}^{(vac)}\Big|_{z\rightarrow 0}\sim \log(z)\hspace{1mm},\hspace{2mm}e^{2\omega^{(vac)}}\Big|_{z\rightarrow 0}\sim \frac{1}{z^{\frac{14}{3}}}.
\end{eqnarray}

Comparing (\ref{bh47}) and (\ref{bh48}) we note that the leading order behaviour of both the black hole and the vacuum solution is identical near the boundary. Hence, the UV central charge (\ref{cc}) for black hole phase will be identical to that with the vacuum solution as mentioned previously section (\ref{S5}). We will explore more about the central charge in the next section.
\section{Thermodynamics of 2D black holes}\label{SBH1}
In the present section, we investigate the thermal properties of 2D black holes (\ref{2dbh}). In particular, we discuss the Wald entropy \cite{Wald:1993nt} of a black hole and interpret its divergences near the black hole horizon. Finally, we compute the Cardy formula \cite{Cardy:1986ie} for  2D black holes and compare it with the Wald entropy to estimate the charge of the corresponding Virasoro generator.

To start with, we calculate the Hawking temperature \cite{Hawking:1974sw} for the 2D black hole (\ref{2dbh})
\begin{align}\label{ht}
   T_H=\frac{1}{2\pi}\sqrt{-\frac{1}{4}g^{tt}g^{\rho\rho}(\partial_\rho g_{tt})^2}\Bigg|_{\rho\rightarrow\sqrt{\mu}}=\hspace{2mm}\frac{\sqrt{\mu}}{\pi}\big(1+6\kappa-\kappa(36d_2+d_8)\big),
\end{align}
where $\sqrt{\mu}$ is the location of the horizon. Notice that, in arriving at (\ref{ht}), we set $Q=\sqrt{\frac{4\mu d_2}{3\log(\mu)}}$, $d_5=0$ and $\mu<<1$ such that hawking temperature reduces to \cite{Lala:2020lge} in the limit $\kappa\rightarrow 0$.
\subsection{Wald entropy }\label{divw}
The  Wald entropy \cite{Wald:1993nt}-\cite{Pedraza:2021cvx} is defined as
\begin{eqnarray}\label{weed}
S_W=-2\pi Y^{abcd}\epsilon_{ab}\epsilon_{cd}\hspace{2mm},\hspace{2mm}Y^{abcd}=\frac{\partial \mathcal{L}}{\partial R_{abcd}},
\end{eqnarray}
where $\mathcal{L}$ is the Lagrangian density\footnote{we have used the notation  $S=\int d^2x\sqrt{-g}\mathcal{L}$ }, $R_{abcd}$ is the Riemann curvature tensor and $\epsilon_{ab}$ is the anti-symmetric tensor with the normalisation condition, $\epsilon^{ab}\epsilon_{ab}=-2$. 

Using (\ref{weed}), one can estimate the Wald entropy for the action (\ref{action 2D})
\begin{eqnarray}\label{we1}
S_W=4\pi\phi-16\kappa\pi\phi e^{-6\omega}\frac{\partial \rho}{\partial z}\frac{d}{d\rho}\Big\{\frac{\partial \rho}{\partial z}\Big(\frac{\partial \omega}{\partial \rho}\Big)\Big\}.
\end{eqnarray}

One can expand the above expression (\ref{we1}) explicitly using the equations (\ref{bh0}), (\ref{bh23}) and (\ref{bh46}) up to leading order in $\xi$ and $\kappa$ as

\begin{align}\label{wee}
S_W=&4\pi\Bigg(\frac{\rho}{6}-\frac{9\kappa\rho}{(-\rho^2+\mu)^2}+\frac{1}{48\mu^{\frac{3}{2}}}\xi\Big\{3\Big(4\rho\sqrt{\mu}+2\mu+(-3\rho^2+\mu)\log\Big(1-\frac{\rho^2}{\mu}\Big)+\nonumber\\
&6\rho^2\log\Big(1-\frac{\rho}{\sqrt{\mu}}\Big)-8\rho\sqrt{\mu}\log\Big(1-\frac{\rho}{\sqrt{\mu}}\Big)-\mu \log\Big(1-\frac{\rho}{\sqrt{\mu}}\Big)-\mu \log\Big(1+\frac{\rho}{\sqrt{\mu}}\Big)+\nonumber\\
&4\rho\sqrt{\mu}\log(\rho)+\tanh^{-1}\Big(\frac{\rho}{\sqrt{\mu}}\Big)(6\rho^2-8\rho\sqrt{\mu}+4(-\rho^2+\mu)\log(\rho))+4\rho\sqrt{\mu}\log(\rho^2-\mu)\nonumber\\
&-(-\rho^2+\mu)\Big(-\text{PolyLog}\Big[2,\frac{\rho^2}{\mu}\Big]+\text{4PolyLog}\Big[2,\frac{\rho}{\mu}\Big]\Big)\Big)Q^2+8\rho^2\mu d_1+4\mu\Big(2\rho\sqrt{\mu}+\nonumber\\
&2\rho^2\tanh^{-1}\Big[\frac{\rho}{\sqrt{\mu}}\Big]+\mu \log(-\sqrt{\mu}+\rho)-\mu \log(\rho+\sqrt{\mu})\Big)d_2+48\mu^{\frac{3}{2}}(d_3+\rho d_4)\Big\}+\nonumber\\
&\kappa\Big\{\frac{3}{4\pi}\Big(\rho\Big(4\log(-\sqrt{\mu}+\rho)+\log\Big(1-\frac{\rho}{\sqrt{\mu}}\Big)^2+2\log\Big(1-\frac{\rho}{\sqrt{\mu}}\Big)\log\Big(\frac{1}{4}\Big(1+\frac{\rho}{\mu}\Big)\Big)\nonumber\\
&-\log\Big(1+\frac{\rho}{\sqrt{\mu}}\Big)^2-4\log(\sqrt{\mu}+\rho)\Big)-4\Big(\sqrt{\mu}-\rho\tanh^{-1}\Big(\frac{\rho}{\sqrt{\mu}}\Big)\Big)\log(-\rho^2+\mu)+\nonumber\\
&4\rho\text{PolyLog}\Big(2,\frac{1}{2}-\frac{\rho}{2\sqrt{\mu}}\Big)\Big)Q^2-d_{12}+\frac{\rho}{\sqrt{\mu}}\Big(d_{11}+\tanh^{-1}\Big[\frac{\rho}{\sqrt{\mu}}\Big]d_{12}\Big)\Big\}\Bigg).
\end{align}

It is evident from the above expression (\ref{wee}) that the Wald entropy  diverges in the near horizon limit i.e. $\rho\rightarrow\sqrt{\mu}$. As we explain below, these divergences are due to the short range correlations between quantum modes across the horizon. A careful inspection, further reveals that these divergences are sourced due to the presence of the higher derivative interaction terms in (\ref{action 2D}).

Below, we explain more about this with the help of a toy model calculation.

$\bullet$ \textbf{A toy model calculation :}

Consider a massive scalar field ($\Phi$) in the black hole background (\ref{2dbh}) that satisfies the Klein-Gordan equation 
\begin{eqnarray}\label{kg1}
(\bigtriangledown^2-m^2)\Phi=0.
\end{eqnarray}

We demand that $\Phi$ satisfies the ``brick wall'' boundary condition i.e. $\Phi=0$ at $z=z_*$, where $z_*$ is the location of black hole horizon. This calculation is analogous to the ’t Hooft’s brick wall model as discussed in \cite{thooft}-\cite{Brustein:2010ms}.

In the black hole background (\ref{2dbh}), equation (\ref{kg1}) takes the form
\begin{eqnarray}\label{kg2}
-\frac{1}{e^{2\omega}}\partial^2_t\Phi+\frac{1}{e^{2\omega}}\partial^2_z\Phi-m^2\Phi=0.
\end{eqnarray}

One can solve the above equation (\ref{kg2}) using method of seperation of variables. We consider $\Phi=e^{iEt}f(z)$, and plug it back into (\ref{kg2}) which yields
\begin{eqnarray}\label{kg3}
\frac{1}{e^{2\omega}}\big(E^2f(z)+\partial^2_zf(z)\big)-m^2f(z)=0.
\end{eqnarray}

In order to proceed further, we substitute $f(z)=\tilde{\rho}(z)e^{iS(z)}$, where $\tilde{\rho}(z)$ is a slowly varying function in $z$ and $S(z)$ is the wildly oscillating phase. On plugging $f(z)$ into (\ref{kg3}), we get
\begin{eqnarray}\label{kg4}
f(z)=\tilde{\rho}(z)e^{\pm i\int dz\sqrt{E^2-e^{2\omega}m^2}}.
\end{eqnarray}

Now we impose an additional boundary condition\footnote{This is called the  Dirichlet boundary condition and the coordinate $\overline{z}$ is located far away from the horizon \cite{thooft}.} on $\Phi$ i.e $\Phi=0$ at $z=\overline{z}$ such that the integral in (\ref{kg4}) become discrete
\begin{eqnarray}\label{ds1}
\int_{z_*}^{\overline{z}} dz\sqrt{E^2-e^{2\omega}m^2}=n(E),
\end{eqnarray}
where $n(E)$ is the density of states which measures the total number of states having energy $E$.

Using (\ref{ze}), we can express the density of states (\ref{ds1}) in terms of $\rho$ as
\begin{eqnarray}\label{ds2}
n(E)=\int_{\sqrt{\mu}}^{\overline{\rho}}d\rho\sqrt{\frac{E^2}{4(\mu-\rho^2)^2}-\frac{e^{2\omega}m^2}{4(\mu-\rho^2)^2}}.
\end{eqnarray}

In principle, one can use the above expression (\ref{ds2}), to estimate the corresponding free energy (F) and entropy (S) for the scalar field ($\Phi$)  as 
\begin{align}\label{FS}
    F=\int_0^{\infty}\frac{n(E)}{1-e^{\beta E}} dE\hspace{2mm}\text{ and }\hspace{2mm}S=\beta^2\partial_{\beta}F\Big|_{\beta=\beta_H},
\end{align}
where $\beta$ is the inverse temperature and $\beta_H$ is the inverse Hawking temperature. 

It is evident from (\ref{ds2}), that the integrand blows up at lower limit i.e. $\rho=\sqrt{\mu}$. This means that the density of states for $\Phi$ diverges near the horizon which leads to divergences in the free energy and entropy (\ref{FS}). In order to get rid of such divergences, we shift the horizon location by an infinitesimal amount $\delta$ i.e. $\rho\rightarrow\sqrt{\mu}+\delta$, where $\delta<<\sqrt{\mu}$. On plugging the shifted horizon back into (\ref{ds2}), we get a finite answer both for the density of states as well as for the entropy \cite{Solodukhin:2011gn,Brustein:2010ms}. This toy model calculation for $\Phi$ gives us an important clue about the interpretation of the above divergences in the Wald entropy (\ref{wee}).

 Recall that, we formulate the action (\ref{action 2D}) by adding matter field content\footnote{By matter field content, we means 2-derivative and 4-derivative interaction terms.} to the pure JT gravity model. Addition of matter field content introduces new degrees of freedom in our theory (\ref{action 2D}), which is analogous to the scalar field ($\Phi$) in the above calculation. Therefore, the divergence in the Wald entropy (\ref{wee}) (that arises due to the addition of the matter field content) is analogous to the divergence in the density of states (\ref{ds2}) for $\Phi$ near the horizon. Therefore, following the above discussion, one can get rid of divergences in the Wald entropy (\ref{wee}) by shifting the actual location of the horizon by an infinitesimal amount namely, $\rho\rightarrow\sqrt{\mu}+\delta$, where $\delta<<\sqrt{\mu}$.

\subsection{Cardy formula for 2D black holes }\label{ccs1}
In literature, there exists an elegant way for counting the number of degrees of freedom associated with 2D CFT. This goes under the name of the Cardy formula \cite{Cardy:1986ie} which is given by 
\begin{eqnarray}\label{s1}
S_C=2\pi\sqrt{\frac{c\Delta}{6}},
\end{eqnarray}
where c is the central charge of the 2D CFT and $\Delta$ is the eigen value of the Virasoro operator $L_0$\footnote{See Appendix \ref{cardy} for a brief discussion on the Cardy formula.}.

It has been found in \cite{Strominger:1997eq} that the entropy computed using the Cardy formula (\ref{s1}) matches with the black hole entropy in the bulk. In particular, the author of \cite{Strominger:1997eq} considers the three-dimensional theory of gravity coupled with matter fields 
\begin{align}\label{ath1}
    S=\frac{1}{16\pi G}\int d^3x\sqrt{-g}\Big(R+\frac{2}{l^2}\Big)+S_m,
\end{align}
where $S_m$ contains the matter field. 

Next, the author computes the central charge for the boundary theory corresponding to (\ref{ath1}) and estimates the boundary degrees of freedom using Cardy formula (\ref{s1}). Remarkably, these boundary degrees of freedom precisely match with the Bekenstein- Hawking entropy of the black holes\footnote{In literature, these black holes are called the BTZ black holes \cite{Banados:1992gq}.} corresponding to (\ref{ath1}). Finally, the author claims that this result holds for any consistent theory of quantum gravity. 

In the present section, we carry out an analysis that is similar in spirit as mentioned above. A similar analysis has been performed by authors in \cite{Castro:2008ms}. In particular, they consider the 2D Einstein-dilaton gravity in the presence of $U(1)$ gauge fields and  compute the central charge associated with the corresponding boundary theory. They determine the boundary degrees of freedom  using Cardy formula and show that it precisely matches with the corresponding Bekenstein- Hawking entropy formula for 2D black holes.

Following similar spirit, our goal is to determine the eigen value of the dilatation operator $L_0$. In order to find $\Delta$, we utilize the fact that the boundary degrees of freedom for the ground state is equivalent to the black hole entropy. Therefore, we compare the Wald entropy of a 2D black hole (\ref{we1}) with the Cardy formula  (\ref{s1}), which yields
\begin{align}
\Delta=\frac{3}{2}\frac{S_W^2}{c\pi^2}.
\end{align}

As our analysis reveals, $\Delta$  receives corrections both due to the presence of 2-derivative and 4-derivative interaction terms in (\ref{action 2D}). We re-scale  $\Delta$ $\rightarrow$ $\tilde{\Delta}=\frac{\Delta}{w}$ ( with $w=3.53\times10^{-5}\mu \kappa^{1.5}$) which finally yields the eigen value for the UV CFT as
\begin{align}\label{s3}
\Delta=\tilde{\Delta}\big|_{UV}=&\frac{c}{24}+\xi F_1+\kappa F_2,
\end{align}
where $F_1$ and $F_2$ are the corrections due to 2-derivative and 4-derivative interactions present in (\ref{action 2D})
\begin{align}\label{s4}
F_1=&\frac{ \kappa^{0.75} }{w}\Big\{\Big(0.0044587-0.0029725\log\Big(-\frac{\delta}{\sqrt{\mu}}\Big)+0.0029725\log(\delta\sqrt{\mu})+\nonumber\\
&0.00148625\log(\mu)\Big)Q^2+\Big(0.00198166+0.000990832\log(\delta)\nonumber\\
&-0.000990832\log\Big(-\frac{\delta}{\sqrt{\mu}}\Big)-0.000495416\log(\mu)\Big)\frac{3\log(\mu)}{4}Q^2+0.01189\mu d_4\Big\},\nonumber\\
F_2=&\frac{\kappa^{0.75}}{w}\Big\{-\frac{0.0267525}{\delta^2}-\frac{0.0535049}{\delta\sqrt{\mu}}+\Big(-0.00428444+0.03567\log(\delta)-\nonumber\\
&0.0123623\log\Big(-\frac{\delta}{\sqrt{\mu}}\Big)+0.00891749\log\Big(-\frac{\delta}{\sqrt{\mu}}\Big)^2-0.03567\log(2\sqrt{\mu})-\nonumber\\
&0.03567\log(-2\delta\sqrt{\mu})+0.03567\tanh^{-1}\Big(1+\frac{\delta}{\sqrt{\mu}}\Big)\log(-2\delta\sqrt{\mu})+\nonumber\\
&0.03567\text{PolyLog}\Big[2,-\frac{\delta}{2\sqrt{\mu}}\Big]\Big)Q^2+0.01189\sqrt{\mu}d_{11}+\Big(-0.01189\sqrt{\mu}+\nonumber\\
&0.01189\sqrt{\mu}\tanh^{-1}\Big(1+\frac{\delta}{\sqrt{\mu}}\Big)\Big)d_{12}\Big\}.
\end{align}

$\bullet$ Note : In arriving at (\ref{s4}), we write the full solution of the gauge field (\ref{e3}) and the dilaton (\ref{e2})
 $$
      A_t^{bh}=A_{t(0)}^{bh}+\frac{\kappa}{\xi}A_{t(1)}^{bh},\hspace{2mm}
      \phi^{bh}=\phi_{(0)}^{bh}+\xi\phi_{(1)}^{bh}+\kappa\phi_{(2)}^{bh}.
$$
  In the near horizon limit, we absorb the integration constant $d_3$ in $\phi_{(1)}^{bh}$ (\ref{bh23}) into the constant $d_{11}$ in $\phi_{(2)}^{bh}$ (\ref{bh46}) without any loss of generality. Similarly, we absorb the additive constant $d_1$ in $A_{t(0)}^{bh}$ (\ref{bh21}) into the constant $d_6$ in $A_{t(1)}^{bh}$ (\ref{bh4a}). Furthermore, we write the constant $d_2$ in terms of the charge $Q$ using (\ref{ht}).

Finally, we have used the fact that $\delta<<\sqrt{\mu}$ and retain terms up to leading order in the couplings $\xi$ and $\kappa$.

\section{Near horizon CFT}\label{SS8}
We now explore the near horizon modes of the theory (\ref{action 2D}). In particular, we look the evidence of a CFT in the near horizon limit and calculate the central charge associated with it.

We start by computing the trace of the stress tensor in the near horizon limit
\begin{eqnarray}\label{NHT1}
g^{\mu\nu}T_{\mu\nu}=\frac{1}{\sqrt{-g}}g^{\mu\nu}\frac{\delta S_{EH}}{\delta g^{\mu\nu}}.
\end{eqnarray}

One can schematically express the above expression (\ref{NHT1}) as 
\begin{eqnarray}\label{NHTS}
g^{\mu\nu}T_{\mu\nu}=T_0+\xi T_1+\kappa T_2,
\end{eqnarray}
where $T_0$ is the trace of the stress tensor for the pure JT gravity theory. On the other hand, $T_1$ and $T_2$ are the correction terms due to the presence of 2-derivative and 4-derivative interactions in (\ref{action 2D}).  

The trace of the stress tensor in the JT gravity is given by
\begin{eqnarray}\label{NHT2}
T_0=e^{-2\omega_0}(\phi_0''-12\phi_0e^{2\omega_0}),
\end{eqnarray}
which turns out to be zero by virtue of equations of motion (\ref{v01}) and (\ref{v02}).

On the other hand, the first order correction in (\ref{NHTS}) due to the presence of 2-derivative interactions is given by
\begin{align}\label{NHT3}
T_1=e^{-2\omega_0}\Big[-12(e^{2\omega_0}\phi_1+2\omega_1\phi_0e^{2\omega_0})+\phi_1''+\frac{\phi_0}{2}A_{t(0)}'^2e^{-2\omega_0}\Big]+2\omega_1e^{-2\omega_0}[-\phi_0''+12\phi_0e^{2\omega_0}],
\end{align}
which  vanishes identically by virtue of (\ref{bh0}) and (\ref{bh23}). 

Finally, we calculate the correction due to the presence of 4-derivative interactions
\begin{align}\label{NHT5}
T_2=-e^{-2\omega_0}\Big[12e^{2\omega_0}(\phi_2+4\phi_0\omega_2)-(\phi_2''+2\omega_2\phi_0'')+A(z)\Big]+4\omega_2e^{-2\omega_0}\Big[-\phi_0''+12\phi_0e^{2\omega_0}\Big],
\end{align}
which also vanishes identically due to equations (\ref{bh0}), (\ref{ze}) and (\ref{bh43}).

Combining (\ref{NHT2})-(\ref{NHT5}), we conclude that the trace of the stress tensor (\ref{NHT1}) vanishes identically in the near horizon limit. These calculations suggest that there exists a conformal field theory in the near horizon limit. Our next step would be to compute the central charge corresponding to this conformal field theory. 

In order to simplify our analysis, we switch off 4-derivative interactions\footnote{See Appendix \ref{vp} for the correction due to 4-derivative interactions.} for the moment and transform the Einstein-Hilbert action (\ref{action 2D}) into the Liouville theory using the following field redefinition \cite{Solodukhin:1998tc}-\cite{Alishahiha:2008tv}
\begin{align}\label{NHT6}
   \phi=\Phi^2=q\Phi_H\psi \hspace{2mm}\text{and}\hspace{2mm} g_{\mu\nu}\rightarrow e^{\frac{2\psi}{q\Phi_H}}g_{\mu\nu},
\end{align}
where q is a constant and $\Phi_H=\Phi|_{horizon}$\footnote{Notice that, we have taken out a common factor $\frac{1}{4}$ in (\ref{NHT7}) in order to be consistent with \cite{Solodukhin:1998tc}.}.

We plug (\ref{NHT6}) into (\ref{action 2D}) which yields
\begin{align}\label{NHT7}
    S=\int d^2x\sqrt{-g}\Big[\frac{1}{4}q\Phi_H\psi R+\frac{1}{2}(\bigtriangledown_{\mu}\psi)^2+3q\Phi_He^{\frac{2\psi}{q\Phi_H}}\psi-\frac{\xi}{16}q\Phi_H\psi e^{-\frac{2\psi}{q\Phi_H}}F^2\Big].
\end{align}

Next, we integrate out the gauge degrees of freedom in the action (\ref{NHT7}) which by virtue of the equation of motion (\ref{v23}), yields
\begin{align}\label{NHT8}
    S_L=\int d^2x\sqrt{-g}\Big[\frac{1}{2}(\bigtriangledown_{\mu}\psi)^2+\frac{1}{4}q\Phi_H\psi R+V(\psi)\Big]\hspace{2mm},\hspace{2mm}
    V(\psi)=3q\Phi_He^{\frac{2\psi}{q\Phi_H}}\psi+\frac{\xi}{8}\frac{ e^{-\frac{2\psi}{q\Phi_H}}}{q\Phi_H\psi}b^2
\end{align}
where $V(\psi)$ is the potential\footnote{See Appendix \ref{vp} for the properties of the potential $V(\psi)$.} of the ``generalised" Liouville theory that contains the 2-derivative interaction term and $b$ is the integration constant. We discuss more about the generalised Liouville theory in the section [\ref{GLT}].

On varying (\ref{NHT8}) with respect to $g_{\mu\nu}$, we obtain the equation of motion for the metric as 
\begin{align}\label{NHT9}
    \frac{1}{2}(\partial_{\mu}\psi)(\partial_{\nu}\psi)-\frac{1}{4}g_{\mu\nu}(\bigtriangledown\psi)^2+\frac{q\Phi_H}{4}(g_{\mu\nu}\square\psi-\bigtriangledown_{\mu}\bigtriangledown_{\nu}\psi)-\frac{1}{2}g_{\mu\nu}V(\psi)=0.
\end{align}

We prefer to solve (\ref{NHT9}) in the following static gauge \cite{Solodukhin:1998tc}
\begin{align}\label{nhg}
    ds^2=-g(x)dt^2+\frac{dx^2}{g(x)}\hspace{2mm},\hspace{2mm}g(x)=\frac{2}{\beta_H}(x-x_H)+O(x-x_H)^2,
\end{align}
where the horizon is located at $x=x_H$.

In the near horizon limit, it is convenient to carry out an analysis in $(t,z)$ coordinate, where $z$ is given by 
\begin{align}
   z=\frac{\beta_H}{2}\log[x-x_H].
    \end{align}
    
In $(t,z)$ coordinates, (\ref{nhg}) reduces to
\begin{align}\label{nhgn}
    ds^2=-g(z)dt^2+g(z)dz^2\hspace{2mm},\hspace{2mm}g(z)=\frac{2}{\beta_H}e^{\frac{2z}{\beta_H}},
\end{align}
where the horizon is located at $z\rightarrow-\infty$.

Next, we note down the components of the stress tensor (\ref{NHT1}) of (\ref{NHT8}) in the gauge (\ref{nhgn}) 
\begin{align}
    T_{tt}&=\frac{1}{4}\Big[(\partial_t\psi)^2+(\partial_z\psi)^2\Big]-\frac{q\Phi_H}{4}\Big[\partial^2_z\psi-\frac{1}{\beta_H}\partial_z\psi\Big]+\frac{1}{2}g(z)V(\psi)\label{NHT10},\\
    T_{tz}&=\frac{1}{2}\partial_t\psi\partial_z\psi-\frac{q\Phi_H}{4}\Big[\partial_z\partial_t\psi-\frac{1}{\beta_H}\partial_t\psi\Big]\label{NHT11}.
\end{align}

Finally, we define the Virasoro generators \cite{Solodukhin:1998tc} in terms of the components of the stress tensor (\ref{NHT10})-(\ref{NHT11}) as 
\begin{align}\label{NHT12}
    L_n=\frac{L}{2\pi}\int _{-\frac{L}{2}}^{\frac{L}{2}}dze^{i\frac{2\pi}{L}nz}T_{++}(z),
\end{align}
where,  $T_{++}=T_{tt}+T_{tz}$ and the integration is on the circle of circumference $L$. At the end of the calculation, we stretch $L$ upto infinity.

Using (\ref{NHT10})-(\ref{NHT11}), in the near horizon limit i.e. $z\rightarrow-\infty$, we obtain
\begin{align}\label{NHT13}
    T_{++}=\frac{1}{4}\Big[(\partial_t+\partial_z)\psi\Big]^2-\frac{q\Phi_H}{4}\Big[\partial_z(\partial_z+\partial_t)\psi-\frac{1}{\beta_H}(\partial_z+\partial_t)\psi\Big].
\end{align}
Notice that, the expression of $T_{++}$ (\ref{NHT13}) does not depend on the form of the potential $V(\psi)$ in the near horizon limit. 

A straightforward calculation reveals that the Virasoro generators (\ref{NHT12}) along with (\ref{NHT13}) satisfy the following commutation relation
\begin{align}
    i\{L_k,L_n\}=(k-n)L_{n+k}+\frac{c_{H}}{12}k\Big(k^2+\Big(\frac{L}{2\pi\beta_H}\Big)^2\Big)\delta_{n+k},0,
\end{align}
where $c_H=3\pi q^2\Phi_H^2$ is the central charge associated with the conformal field theory near the horizon.

Using (\ref{bh0}) and (\ref{bh23}), one can further rewrite the central charge as
\begin{align}\label{cch}
    c_H=&\hspace{1mm}3\pi q^2\Bigg[\frac{\sqrt{\mu}}{6}+\xi\Bigg\{\Bigg\{\frac{3 }{8\sqrt{\mu}}-\frac{
    \tanh^{-1}\Big[\frac{\delta+\sqrt{\mu}}{\sqrt{\mu}}\Big]}{8\sqrt{\mu}}-\frac{\log[2]}{16\sqrt{\mu}}-\frac{\log\Big[1-\frac{(\delta+\sqrt{\mu})^2}{\mu}\Big]}{8\sqrt{\mu}}+\frac{\log[\sqrt{\mu}]}{4\sqrt{\mu}}\nonumber\\
    &-\frac{3\log\Big[1-\frac{\delta+\sqrt{\mu}}{\sqrt{\mu}}\Big]}{16\sqrt{\mu}}+\frac{\log[(\delta+\sqrt{\mu})^2-\mu]}{4\sqrt{\mu}}\Bigg\}Q^2+\Big\{\frac{1}{6}\sqrt{\mu}+\frac{1}{6}\sqrt{\mu}\tanh^{-1}\Big[\frac{\delta+\sqrt{\mu}}{\sqrt{\mu}}\Big]+\nonumber\\
    &\frac{1}{12}\sqrt{\mu}\log[\delta]-\frac{1}{12}\sqrt{\mu}\log[2\sqrt{\mu}]\Big\}\frac{3\log(\mu)}{4\mu}Q^2+\sqrt{\mu} d_4\Bigg\}\Bigg],
\end{align}
where the constant $d_1$ is absorbed in $d_{6}$ and $d_3$ in $d_{11}$ as discussed in (\ref{s4}).

Notice that, the expression (\ref{cch}) diverges near the horizon which is due to the divergences in the corresponding density of states as we have discussed in the Section (\ref{divw}). Therefore, in order to obtain a finite answer, we calculate the central charge in the limit $\rho\rightarrow\sqrt{\mu}+\delta$, where $\delta<<\sqrt{\mu}$.

\section{Generalised Liouville Theory and Weyl anomaly}\label{GLT}
In this section, we study the generalised Liouville theory in 2D that contains the 2-derivative interaction terms (\ref{NHT8}). In particular, we focus on the Weyl transformation properties of the generalised Liouville theory and the Weyl anomaly associated with it. 

Liouville theory is a conformal field theory in 2D which is dual to the Einstein gravity with negative cosmological constant in three-dimensions \cite{Coussaert:1995zp}-\cite{Grumiller:2007wb}. The action for the Liouville theory in two-dimensions is given by 
\begin{align}\label{sl1}
    S_L=\int d^2x\sqrt{-g}\Big[\frac{1}{2}g^{\mu\nu}\bigtriangledown_{\mu}\psi\bigtriangledown_{\nu}\psi+\frac{R}{\beta}\psi-\frac{m^2}{\beta^2}e^{\beta\psi}\Big],
\end{align}
where $R$ is the Ricci scalar, $\psi$ is the scalar field and ($\beta$, $m$) are the constants.

It is shown in \cite{Grumiller:2007wb} that one can construct the Liouville theory (\ref{sl1}) by consistent dimensional reduction of pure Einstein-Hilbert action in $D$ dimensions. The authors in \cite{Grumiller:2007wb} start with the following action
\begin{align}\label{sl2}
    S_D=\int d^Dx\sqrt{-g_{(D)}}R^{(D)}.
\end{align}

The dimensional reduction ansatz for space-time metric is given by
\begin{align}\label{sl3}
    ds^2=g_{\mu\nu}^{(D)}dx^{\mu}dx^{\nu}=g_{\alpha\beta}dx^{\alpha}dx^{\beta}+\frac{1}{\lambda}\phi^{\frac{2}{(D-2)}}d\Omega^2_{S_{D-2}},
    \end{align}
where ($\mu,\nu$) are $D$ dimensional indices, ($\alpha,\beta$) are 2 dimensional indices and $\lambda$ is the parameter having dimensions $[L]^{-2}$. Next, the authors parameterize the dimensions by $D=2+\epsilon$ and plug (\ref{sl3}) into (\ref{sl2}) to obtain the action for Liouville theory in the limit $\epsilon\rightarrow0$.

The stress energy tensor and the equation of motion for the field $\psi$ corresponding to the Liouville theory (\ref{sl1}) are  given by 
\begin{align}
T_{\mu\nu}&=\frac{1}{\sqrt{-g}}\frac{\partial S_{L}}{\partial g^{\mu\nu}}\nonumber\\
&=\frac{1}{2}\bigtriangledown_{\mu}\psi\bigtriangledown_{\nu}\psi-\frac{1}{4}g_{\mu\nu}(\bigtriangledown\psi)^2+\frac{1}{\beta}(g_{\mu\nu}\bigtriangledown^2\psi-\bigtriangledown_{\mu}\bigtriangledown_{\nu}\psi)+\frac{m^2}{2\beta^2}g_{\mu\nu}e^{\beta\psi},\label{sess1}\\
\bigtriangledown^2\psi&=\frac{R}{\beta}-\frac{m^2}{\beta}e^{\beta\psi}.\label{sess2}
\end{align}

Using (\ref{sess2}), one can compute the trace of stress tensor (\ref{sess1}) as
\begin{align}\label{trn}
   <T^{\mu}_{\mu}>= \frac{R}{\beta^2},
\end{align}
which does not vanish in curved space-time. This is what is known as the Weyl anomaly, where the coefficient $\frac{1}{\beta^2}$ is related to the central charge of the CFT.

Next, we look at the Weyl transformation properties of the Liouville theory (\ref{sl1}). In order to proceed, we consider the following field transformations \cite{Jackiw:2005su}
\begin{align}\label{sl4}
    g_{\mu\nu}\rightarrow e^{2\sigma}g_{\mu\nu}\hspace{2mm},\hspace{2mm}\psi\rightarrow\psi-\frac{2}{\beta}\sigma,
\end{align}
where $\sigma\equiv\sigma(t,z)$. Under the above transformation (\ref{sl4}), the action (\ref{sl1}) is transformed (up to boundary terms) as, $S_L\rightarrow S_L+\delta S_L$ where the difference is denoted as 
\begin{align}\label{sl5}
   \delta S_L=-\frac{2}{\beta^2}\int d^2x\sqrt{-g}\Big[R\sigma+g^{\mu\nu}\bigtriangledown_{\mu}\sigma\bigtriangledown_{\nu}\sigma\Big].
\end{align}

Equation (\ref{sl5}) suggests that the Liouville theory (\ref{sl1}) is not invariant under the transformation\footnote{See \cite{Jackiw:2005su} to obtain the Liouville theory in $D=2$ dimensions from Weyl invariant theories in $D>2$. } (\ref{sl4}).
However, the difference  $\delta S_L$ (\ref{sl5}) does not depend on the field $\psi$. As a result, the  equation of motion for $\psi$ (\ref{sess2})
remains invariant under the transformation (\ref{sl4}).

Notice that, the difference $\delta S_L$ (\ref{sl5}) can be set equal to zero (up to boundary terms) if we impose the equation of motion for $\sigma$
\begin{align}
    \bigtriangledown_{\mu}\bigtriangledown^{\mu}\sigma=R.\label{sric1}
\end{align}

One can solve (\ref{sric1}) for $\sigma$ in the  static light cone gauge (\ref{sg}) which yields a solution of the form
\begin{align}
 \sigma=-2\omega+z b_1+b_2,   
\end{align}
where $b_1$ and $b_2$ are the integration constants. Therefore, given the onshell condition (\ref{sric1}), the action (\ref{sl1}) is claimed to be invariant under the Weyl re-scaling (\ref{sl4}).

The Liouville theory (\ref{NHT8}) that we obtain is different from the standard Liouville theory (\ref{sl1}) in the sense that (\ref{NHT8}) does not reduce to (\ref{sl1}) in the limit $\xi\rightarrow0$. We are interested to look at the transformation properties of this generalised Liouville theory (\ref{NHT8}) under the following field redefinition 
\begin{align}\label{sft1}
     g_{\mu\nu}\rightarrow e^{2\sigma}g_{\mu\nu}\hspace{2mm},\hspace{2mm}\psi\rightarrow\psi-\tilde{c}_H\sigma\hspace{2mm},\hspace{2mm}\text{where}\hspace{2mm}\tilde{c}_H=\sqrt{\frac{c_H}{3\pi}}.
\end{align}

Under the above transformation (\ref{sft1}), the action (\ref{NHT8}) is transformed (upto boundary terms) as 
\begin{align}\label{stransact}
     \tilde{S_L}=&\int d^2x\sqrt{-g}\Big[\frac{1}{2}(\bigtriangledown_{\mu}\psi)^2+\frac{1}{4}\tilde{c}_H\psi R+3\tilde{c}_He^{\frac{2\psi}{\tilde{c}_H}}\psi+\frac{\xi b^2}{8}\frac{ e^{-\frac{2\psi}{\tilde{c}_H}}}{\tilde{c}_H(\psi-\sigma \tilde{c}_H)}e^{4
     \sigma}+\nonumber\\
     &\frac{1}{2}\tilde{c}_H\psi\bigtriangledown_{\mu}\bigtriangledown^{\mu}\sigma-\frac{\sigma}{4}\tilde{c}_H^2R-3\tilde{c}_H^2\sigma e^{\frac{2\psi}{\tilde{c}_H}}\Big],
\end{align}
 which is clearly not invariant. On top of that, even the dynamics of the scalar field $(\psi)$ is influenced deriving the transformation (\ref{sft1}). 
 
The stress energy tensor and the equation of motion for $\psi$ that follows from (\ref{NHT8}) are given by 
\begin{align}
T_{\mu\nu}&=\frac{1}{\sqrt{-g}}\frac{\delta S_{L}}{\delta g^{\mu\nu}}\nonumber\\
&=\frac{1}{2}\bigtriangledown_{\mu}\psi\bigtriangledown_{\nu}\psi-\frac{1}{4}g_{\mu\nu}(\bigtriangledown\psi)^2+\frac{1}{4}\tilde{c}_H(g_{\mu\nu}\bigtriangledown^2\psi-\bigtriangledown_{\mu}\bigtriangledown_{\nu}\psi)-\frac{1}{2}g_{\mu\nu}V(\psi),\label{sess3}\\
\bigtriangledown^2\psi&=\frac{1}{4}\tilde{c}_HR+V'(\psi).\label{sess4}
\end{align}

Using (\ref{sess4}), we compute the trace of stress tensor (\ref{sess3}), which yields
\begin{align}\label{sstn0}
   <T^{\mu}_{\mu}>\sim\frac{c_H}{48\pi}R+O\Big(\frac{1}{\sqrt{c_H}}\Big).
\end{align}
 
 Equation (\ref{sstn0}) confirms that the theory (\ref{NHT8}) is not Weyl Invariant. This is what we identify as the Weyl anomaly \cite{Castro:2019vog} for the generalised Liouville theory (\ref{stransact}). 
 
 After some algebra, one can express the transformed action (\ref{stransact}) as
 \begin{align}\label{svar1}
     S_L\rightarrow S_L+&\int d^2x\sqrt{-g}\Bigg[
     \frac{1}{2}\tilde{c}_H\psi\bigtriangledown_{\mu}\bigtriangledown^{\mu}\sigma-\frac{\sigma}{4}\tilde{c}_H^2R-3\tilde{c}_H^2\sigma e^{\frac{2\psi}{\tilde{c}_H}}+\nonumber\\
     &\frac{\xi b^2}{8}\frac{e^{-\frac{2\psi}{\tilde{c}_H}}}{\tilde{c}_H}\Big\{\frac{e^{4\sigma}}{(\psi-\sigma \tilde{c}_H)}-\frac{1}{\psi}\Big\}\Bigg].
 \end{align}

Notice that, following our previous arguments, the variation $\delta S_L$ (\ref{svar1}) can be set equal to zero if we impose the equation of motion for $\sigma$ 
\begin{align}\label{sem1}
    \frac{1}{2}\tilde{c}_H\psi\bigtriangledown_{\mu}\bigtriangledown^{\mu}\sigma-\frac{\sigma}{4}\tilde{c}_H^2R-3\tilde{c}_H^2\sigma e^{\frac{2\psi}{\tilde{c}_H}}+\frac{\xi b^2}{8}\frac{e^{-\frac{2\psi}{\tilde{c}_H}}}{\tilde{c}_H}\Big\{\frac{e^{4\sigma}}{(\psi-\sigma \tilde{c}_H)}-\frac{1}{\psi}\Big\}=0.
\end{align}

Next, we solve the equation of motion for $\sigma$  (\ref{sem1}) in the static light cone gauge (\ref{sg}). To start with, we perturbatively expand the fields $\psi$, $\omega$ and $\sigma$ treating $\xi$ as an expansion parameter
\begin{align}
    \psi&=\psi_0+\xi \psi_1,\label{sig2}\\
    \omega&=\omega_0+\xi \omega_1,\label{sig3}\\
    \sigma&=\sigma_0+\xi\sigma_1\label{sig4}.
\end{align}

The subscript (0) denotes  the zeroth order fields and the subscript (1) denotes the first order correction in the fields due to the presence of 2-derivative interactions. 

Using (\ref{sig2})-(\ref{sig4}), one can write the zeroth order equation of motion for $g_{\mu\nu}$ (\ref{sess3}), $\psi$ (\ref{sess4}) and $\sigma$ (\ref{sem1}) as
\begin{align}
    \psi_0'^2+\tilde{c}_H(\omega_0'\psi_0'-\psi_0'')+6\tilde{c}_He^{2\big(\frac{\psi_0}{\tilde{c}_H}+\omega_0\big)}\psi_0=0,\label{nee1}\\
       \psi_0'^2+\tilde{c}_H\omega_0'\psi_0'-6\tilde{c}_He^{2\big(\frac{\psi_0}{\tilde{c}_H}+\omega_0\big)}\psi_0=0,\label{nee2}\\
       \psi_0''+\frac{1}{2}\tilde{c}_H\omega_0''-3\psi_0'e^{2\big(\frac{\psi_0}{\tilde{c}_H}+\omega_0\big)}\Big(\tilde{c}_H+2\psi_0\Big)=0,\label{nee3}\\
       \psi_0\sigma_0''+\tilde{c}_H\Big[\omega_0''-6e^{2\big(\frac{\psi_0}{\tilde{c}_H}+\omega_0\big)}\Big]\sigma_0=0.\label{nee4}
\end{align}

Notice that, the above equations (\ref{nee1})-(\ref{nee4}) are the coupled non-linear differential equations and it is difficult to solve them exactly. Therefore, we solve these equations in the large $\tilde{c}_H$ limit and ignore all terms of the order $O(\frac{1}{\tilde{c}_H})$.

Using (\ref{nee1})-(\ref{nee3}), one can write the equation for $\psi_0$ in the large $\tilde{c}_H$ limit as
\begin{align}
    4\psi_0\psi_0'+\tilde{c}_H\psi_0'-\tilde{c}_H=0.\label{nee5}
\end{align}

On solving (\ref{nee5}) for $\psi_0$, we obtain
\begin{align}\label{nee6}
    \psi_0=\frac{1}{4}\Big(-\tilde{c}_H\pm\sqrt{8c_1+\tilde{c}_H\big(8z+\tilde{c}_H\big)}\Big),
\end{align}
where $c_1$ is the integration constant.

Finally, using (\ref{nee1})-(\ref{nee3}) and (\ref{nee6}) in the equation of motion for $\sigma$ (\ref{nee4}),  we obtain 
\begin{align}
   \sigma_0''+f(z)\sigma_0=0,\label{nee7}
\end{align}
where $f(z)$ is given by
\begin{align}\label{nee8}
    f(z)=\frac{32}{\tilde{c}_H^2\Big(\frac{8 c_1}{\tilde{c}_H^2}+\frac{8z}{\tilde{c}_H}+1\Big)^{\frac{3}{2}}\Big(1+\sqrt{\big(\frac{8 c_1}{\tilde{c}_H^2}+\frac{8z}{\tilde{c}_H}+1\big)}\Big)}=\frac{16}{\tilde{c}_H^2}+O\Big(\frac{1}{\tilde{c}_H^3}\Big).
\end{align}

On solving (\ref{nee7}) using (\ref{nee8}) in the large $\tilde{c_H}$ limit, we obtain
\begin{align}
    \sigma_0=c_2\cos{\Big(\frac{4z}{\tilde{c}_H}\Big)}+c_3\sin{\Big(\frac{4z}{\tilde{c}_H}\Big)},
\end{align}
where $c_2$ and $c_3$ are the integration constants.

Finally, we note down equations at leading order in $\xi$ which yield

\begin{align}
    &2\psi_1'\psi_0'+\tilde{c}_H(\omega_0'\psi_1'+\omega_1'\psi_0'-\psi_1'')+2e^{2\omega_0}\Bigg[3\tilde{c}_He^{2\frac{\psi_0}{\tilde{c}_H}}\Big(\frac{2}{\tilde{c}_H}\psi_0\psi_1+\psi_1\Big)+\frac{b^2}{8\psi_0}\frac{1}{\tilde{c}_H}e^{-2\frac{\psi_0}{\tilde{c}_H}}+\nonumber\\
    &6\omega_1\psi_0\tilde{c}_He^{2\frac{\psi_0}{\tilde{c}_H}}\Bigg]=0,\label{nde1}\\
    &2\psi_1'\psi_0'+\tilde{c}_H(\omega_0'\psi_1'+\omega_1'\psi_0')-2e^{2\omega_0}\Bigg[3\tilde{c}_He^{2\frac{\psi_0}{\tilde{c}_H}}\Big(\frac{2}{\tilde{c}_H}\psi_0\psi_1+\psi_1\Big)+\frac{b^2}{8\psi_0}\frac{1}{\tilde{c}_H}e^{-2\frac{\psi_0}{\tilde{c}_H}}+\nonumber\\
    &6\omega_1\psi_0\tilde{c}_He^{2\frac{\psi_0}{\tilde{c}_H}}\Bigg]=0,\label{nde2}\\
    &\psi_1''+\frac{1}{2}\tilde{c}_H\omega_1''-3e^{2\big(\frac{\psi_0}{\tilde{c}_H}+\omega_0\big)}\Bigg[\tilde{c}_H\Big\{\psi_1'+\frac{2}{\tilde{c}_H}\psi_1\psi_0'+2\omega_1\psi_0'\Big\}+2\Big\{\psi_0'\psi_1+\psi_1'\psi_0+\nonumber\\
    &\frac{2}{\tilde{c}_H}\psi_1\psi_0'\psi_0+2\omega_1\psi_0'\psi_0\Big\}\Bigg]+\frac{b^2}{4\psi_0}\frac{1}{\tilde{c}_H}\psi_0'e^{2\big(-\frac{\psi_0}{\tilde{c}_H}+\omega_0\big)}\Bigg(\frac{1}{2\psi_0}+\frac{1}{\tilde{c}_H}\Bigg)=0,\label{nde3}\\
    & \psi_1\sigma_0''+ \psi_0\sigma_1''+\tilde{c}_H\Big[\omega_0''\sigma_1+\omega_1''\sigma_0\Big]-6\tilde{c}_He^{2\big(\frac{\psi_0}{\tilde{c}_H}+\omega_0\big)}\Big(\sigma_1+2\omega_1\sigma_0+\frac{2}{\tilde{c}_H}\psi_1\sigma_0\Big)+\nonumber\\
    &\frac{ b^2}{4}\frac{e^{-2\big(\frac{\psi_0}{\tilde{c}_H}-\omega_0\big)}}{\tilde{c}_H^2}\Bigg[\frac{e^{4\sigma_0}}{(\psi_0-\sigma_0 \tilde{c}_H)}-\frac{1}{\psi_0}\Bigg]=0.\label{nde4}
\end{align}
Obtaining solutions for (\ref{nde1})-(\ref{nde4}) are quite involved which we therefore do not pursue here.

\section{Conclusion}\label{sum}
To summarise, in the present work we extend the notion of 2D Einstein-Maxwell-Dilaton gravity by incorporating the most general form of quartic interactions allowed by the diffeomorphism invariance. We further explore the effects of adding such quartic interactions on the dual field theory observables at strong coupling. Below we outline a couple of future directions along which this work can be further persuaded. 
\begin{itemize}
    \item It is natural to further generalise  our results in the presence of SU(2) Yang-Mills fields and look for its imprints on the holographic stress tensor as well as central charge associated with the boundary theory. It is noteworthy to mention that the SU(2) Yang-Mills fields are responsible for first order phase transition in 2D gravity \cite{Lala:2020lge}. Therefore, it would be an interesting project to explore phase transition in the presence of quartic couplings.

    \item Finally, it would be an interesting project to explore the holographic renormalisation group flow and holographic c-theorem \cite{Myers:2010tj}-\cite{Suh:2020qnl} in the context of 2D gravity theory that contains the most generic quartic interactions (\ref{action 2D}).
\end{itemize}

We hope to be able to report some of these results in the near future.
\section*{Acknowledgments}
The authors are indebted to the authorities of Indian Institute of Technology, Roorkee for their unconditional support towards researches in basic sciences. DR would like to acknowledge The Royal Society, UK for financial
assistance. DR  would also like to acknowledge the Grant (No. SRG/2020/000088)
received from The Science and Engineering Research Board (SERB), India.

\newpage
\appendix
\section{Non-abelian generalization of JT gravity}\label{Non-abelian}
In this Appendix, we will derive the most general 2D action for Einstein-dilaton gravity coupled with U(1) gauge and SU(2) Yang-Mills fields that contains the 4-derivative interaction terms. 

We start with the most general 5D action that contains the 2-derivative interaction terms \begin{align}\label{Na 5D}
S^{(2)}_{5D}=& \int d^5x\sqrt{-g_{(5)}}\Big[\alpha_1(\lambda + R)-\frac{\alpha_2}{4}F^2+\frac{\alpha_3}{3}\epsilon^{MNOPQ}A_MF_{NO}F_{PQ}-\frac{\alpha_4}{4}\big(F^{(a)}\big)^2+\nonumber\\
&\frac{\alpha_5}{4}\epsilon^{MNOPQ}A_MF^{(a)}_{NO}F^{(a)}_{PQ}\Big]
\end{align}
 where $\alpha_i$ $(i=1,..,5)$ are the respective coupling constants. 
 
 Next, we will add the following 4-derivative gauge invariant interaction terms to the above action (\ref{Na 5D})
 
\begin{align}\label{Na4 5D}
S^{(4)}_{5D}=&\int d^5x\sqrt{-g_{(5)}}\Big[\beta_1 R^2+\beta_2 (R_{MN})^2 + \beta_3 (R_{MNOP})^2+\beta_4 RF^2+ \beta_5R^{MN}F_{MO}{F_{N}}^O +\nonumber\\
&\beta_6 R_{MNOP}F^{MN}F^{OP}+\beta_7 R_{MNOP}F^{MO}F^{NP}+\beta_8 (F^2)^2+ \beta_9 F^{MN}F_{NO}F^{OP}F_{PM}+\nonumber\\
&\beta_{10}\bigtriangledown^MF_{MN}\bigtriangledown^O{F_{O}}^N+\beta_{11}\bigtriangledown_MF_{NO}\bigtriangledown^MF^{NO}+\beta_{12}\bigtriangledown_MF_{NO}\bigtriangledown^NF^{MO}+\nonumber\\
&\beta_{13}\bigtriangledown^2F^2+\beta_{14}\bigtriangledown_M\bigtriangledown^NF_{NO}F^{MO}+\beta_{15}\bigtriangledown^N\bigtriangledown_MF_{NO}F^{MO}+\epsilon^{MNOPQ}\big\{F_{MN}\times\nonumber\\
&(\beta_{16}F_{OP}\bigtriangledown^RF_{RQ}+\beta_{17}F_{OR}\bigtriangledown^RF_{PQ}+\beta_{18}F_{OR}\bigtriangledown_PF_{QS}g^{RS})+\beta_{19}A_M R_{NOKL}\times\nonumber\\
&{R_{PQ}}^{KL}\big\}+\beta_{20}\epsilon_{NOPQR}\epsilon^{NIJKL}F^{OP}F^{QR}F_{IJ}F_{KL}+\delta_1R(F^{(a)})^2+\delta_2R^{MN}F^{(a)}_{MO}{F^{(a)}_{N}}^O\nonumber\\
&+\delta_3 R_{MNOP}F^{(a)MN}F^{(a)OP}+\delta_4 R_{MNOP}F^{(a)MO}F^{(a)NP}+\delta_5((F^{(a)})^2)^2+\nonumber\\
&\delta_6F^{(a)}_{MN}F^{(b)MN}F^{(a)}_{OP}F^{(b)OP}+\delta_7F^{(a)MN}F^{(a)}_{NO}F^{(b)OP}F^{(b)}_{PM}+\delta_8F^{(a)MN}F^{(b)}_{NO}F^{(a)OP}F^{(b)}_{PM}+\nonumber\\
&\delta_9F^{(a)MN}F^{(b)}_{NO}F^{(b)OP}F^{(a)}_{PM}+\delta_{10}\bigtriangledown^2(F^{(a)})^2+\delta_{11}\bigtriangledown_M\bigtriangledown^NF^{(a)}_{NO}F^{(a)MO}+\nonumber\\
&\delta_{12}\bigtriangledown^N\bigtriangledown_MF^{(a)}_{NO}F^{(a)MO}+\epsilon_{NOPQR}\epsilon^{NIJKL}\big\{\delta_{13}F^{(a)OP}F^{(a)QR}F^{(b)}_{IJ}F^{(b)}_{KL}+\nonumber\\
&\delta_{14}F^{(a)OP}F^{(b)QR}F^{(a)}_{IJ}F^{(b)}_{KL}\big\}+\delta_{15}F^2(F^{(a)})^2+\delta_{16}\epsilon_{MNORQ}(\bigtriangledown_PF^{PM})F^{(a)NO}F^{(a)RQ}+\nonumber\\
&\delta_{17}F_{MN}F_{OP}F^{(a)MN}F^{(a)OP}+\epsilon^{MOPQR}\epsilon_{MIJKL}\big\{\delta_{18}F_{OP}F_{QR}F^{(a)IJ}F^{(a)KL}+\nonumber\\
&\delta_{19}F_{OP}F^{(a)}_{QR}F^{IJ}F^{(a)KL}\big\}+\delta_{20}\epsilon_{MNORQ}F^{PM}\bigtriangledown_P(F^{(a)NO}F^{(a)RQ})+\nonumber\\
&\delta_{21}\bigtriangledown_P(\epsilon_{MNORQ}F^{PM}F^{(a)NO}F^{(a)RQ})+\delta_{22}F^{(a)RP}F^{(a)}_{SP}F^{SQ}F_{RQ}+\nonumber\\
&\epsilon^{MNOPQ}\big\{\delta_{23}(\bigtriangledown_M{F_N}^R)F^{(a)}_{OR}F^{(a)}_{PQ}+\delta_{24}\bigtriangledown_M(F^{(a)}_{OR}F^{(a)}_{PQ}){F_N}^R+\delta_{25}(\bigtriangledown^RF_{MN})F^{(a)}_{OR}F^{(a)}_{PQ}+\nonumber\\
&\delta_{26}\bigtriangledown^R(F^{(a)}_{OR}F^{(a)}_{PQ})F_{MN}\big\}+\delta_{27}\bigtriangledown_M(\epsilon^{MNOPQ}{F_N}^RF^{(a)}_{OR}F^{(a)}_{PQ})+\nonumber\\
&\delta_{28}\bigtriangledown^R(\epsilon^{MNOPQ}F_{MN}F^{(a)}_{OR}F^{(a)}_{PQ})\Big]
\end{align}
where $\beta_i$ and $\delta_j$ $(i=1,..,20$ and $j=1,..,28)$ are the respective coupling constants.
 
 On adding (\ref{Na 5D}) and (\ref{Na4 5D}), we get the required action that contains all 2-derivative as well as 4-derivative interactions. However, we can eliminate the various interaction terms in (\ref{Na4 5D}) using a proper redefinition of fields.
 
 Consider the following redefinition of fields
 \begin{align}
     g^{RS}\rightarrow g^{RS}+\delta g^{RS},\hspace{2mm}A_N\rightarrow A_N+\delta A_N,\hspace{2mm}A_N^{(a)}\rightarrow A_N^{(a)}+\delta A_N^{(a)}
 \end{align}
 such that the action transform as $S\rightarrow S'=S+\delta S$ where $S=S_{5D}^{(2)}+S_{5D}^{(4)}$ and $S'$ is called the transformed action.
 
 The most general variation of fields that contains 2-derivative interactions takes the form
 \begin{align}
 \delta g^{RS} = & \mu_1R^{RS}+\mu_2F^{RP}{F^S}_P+\mu_3Rg^{RS}+\mu_4F^2g^{RS}+\mu_5g^{RS}+\mu_6(F^{(a)})^2g^{RS}+\nonumber\\
&\mu_7F^{(a)RP}{F^{(a)S}}_P,\label{naa1}\\
\delta A_N= &\lambda_1A_N+\lambda_2\bigtriangledown^MF_{MN}+\lambda_3\epsilon_{NOPQR}F^{OP}F^{QR}+\lambda_4\epsilon_{NOPQR}F^{(a)OP}F^{(a)QR},\label{naa2}\\
\delta A^{(a)}_N= &\sigma_1A^{(a)}_N+\sigma_2\bigtriangledown^MF^{(a)}_{MN}+\sigma_3\epsilon_{NOPQR}F^{(a)OP}F^{QR},\label{naa3}
 \end{align}
 where $\mu_i$, $\lambda_j$ and $\sigma_k$ $(i=1,..,7,j=1,..,4$ and $ k=1,..,3)$ are the respective constants. 
  
  In order to get rid of various interaction terms in (\ref{Na 5D}) and (\ref{Na4 5D}), we transformed the action $S$ (where $S=S_{5D}^{(2)}+S_{5D}^{(4)}$) into $S'$ using (\ref{naa1})-(\ref{naa3}) with the following particular choice of constants 
  \begin{align}
      &\mu_1=-\frac{4\beta_2}{3\alpha_1},\hspace{2mm}\mu_2 = \frac{-2\alpha_2\beta_2}{3\alpha_1^2} ,\hspace{2mm}\mu_3=\frac{4}{9\alpha_1}(2\beta_1+\beta_2), \hspace{2mm}\mu_4=\frac{2\alpha_2}{27\alpha_1^2}(2\beta_2+\beta_1),\hspace{2mm}\mu_5 = -\frac{2}{3},\nonumber\\
&\mu_6 = \frac{2}{27\alpha_1}[9\delta_2+6\delta_1+\frac{\alpha_4}{\alpha_1}(2\beta_2+\beta_1)],\hspace{2mm} \mu_7=\frac{-2}{3\alpha_1}[\delta_2+\frac{\alpha_4}{\alpha_1}\beta_2],\hspace{2mm}\lambda_1=\frac{-1}{3},\nonumber\\
&\lambda_2=\frac{-4\alpha_2}{3\alpha_3\alpha_5}[2\delta_{18}-\frac{\alpha_5}{\alpha_3}\beta_{20}],\hspace{3mm}\lambda_3=\frac{4\beta_{20}}{3\alpha_3},\hspace{3mm} \lambda_4=\frac{1}{3\alpha_3}[2\delta_{18}-\frac{\alpha_5}{\alpha_3}\beta_{20}]
  \end{align}
  and $\sigma_k=0$ \footnote{We choose $\sigma_k=0$ to preserve the gauge invariance.} which yields
  \begin{align}\label{Na4f 5D}
      S'=&\int d^5x\sqrt{-g_{(5)}}\Big[\eta_1\lambda +\eta_2R-\frac{\eta_3}{4}F^2-\frac{\eta_4}{4}\big(F^{(a)}\big)^2+
\frac{\eta_5}{4}\epsilon^{MNOPQ}A_MF^{(a)}_{NO}F^{(a)}_{PQ}+\nonumber\\
&\eta_6[R_{MNOP}]^2+\eta_7(F^2)^2+\eta_8F^{SP}F_{PR}F^{RQ}F_{QS}+\eta_9\bigtriangledown_MF^{MN}\bigtriangledown^OF_{ON}+\nonumber\\
&+\epsilon^{MNOPQ}\big\{\eta_{10}F_{MN}F_{OP}\bigtriangledown^RF_{RQ}+\eta_{11}F_{MN}F_{OR}\bigtriangledown^RF_{PQ}+\eta_{12}F_{MN}F_{OR}\bigtriangledown_P{F_Q}^R+\nonumber\\
&\eta_{13}A_MR_{NOIJ}{R_{PQ}}^{IJ}\big\}+\eta_{14}\big(\big(F^{(a)}\big)^2\big)^2+\eta_{15}F^2\big(F^{(a)}\big)^2+\eta_{16}F^{(a)RP}F^{(a)}_{SP}F^{SQ}F_{RQ}+\nonumber\\
&\eta_{17}F^{(a)RP}F^{(a)}_{PS}F^{(b)SQ}F^{(b)}_{QR}+\eta_{18}\epsilon_{NOPQR}\epsilon^{NIJKL}F^{(a)OP}F^{(a)QR}F^{(b)}_{IJ}F^{(b)}_{KL}+\nonumber\\
&\eta_{19}R_{MNOP}F^{(a)MN}F^{(a)OP}+\eta_{20}R_{MNOP}F^{(a)MO}F^{(a)NP}+\eta_{21}F_{MN}F_{PQ}F^{(a)MN}F^{(a)PQ}+\nonumber\\
&\eta_{22}\epsilon^{MOPQR}\epsilon_{MIJKL}F_{OP}F^{(a)}_{QR}F^{IJ}F^{(a)KL}\Big]
  \end{align}
  where $\eta_i$ $(i=1,..,22)$ are the new coupling constants respectively. 
  
  We can express the new coupling constants $\eta_i$ in terms of the old coupling constants $\alpha_j$, $\beta_k$ and $\delta_l$ as follows 
  \begin{align}
      &\eta_1 =\frac{8}{3}\alpha_1, \hspace {2mm}\eta_2=2\alpha_1-\frac{4\lambda}{9}(5\beta_1+\beta_2),\hspace{2mm}\eta_3=\frac{\lambda\alpha_2}{27\alpha_1}(4\beta_2+20\beta_1)+\frac{2\alpha_2}{3},\nonumber\\
&\eta_4=\frac{8\lambda}{3}\Big(2\delta+\frac{5}{3}\delta_1\Big)+\frac{4\lambda\alpha_4}{27\alpha_1}(5\beta_1+\beta_2)+\frac{4\alpha_4}{3},\hspace{2mm}\eta_5=\frac{2\alpha_5}{3},\hspace{2mm}\eta_6=\frac{4\beta_3}{3},\nonumber\\
&\eta_7=\frac{\alpha_2^2}{108\alpha_1^2}(\beta_1-7\beta_2)-\frac{4\beta_8}{3},\hspace{2mm}\eta_8=\frac{1}{3}\Big(\frac{\alpha_2^2}{\alpha_1^2}\beta_2-4\beta_9\Big),\hspace{2mm}\eta_9=\frac{-4\alpha_2^2}{3\alpha_3\alpha_5}\Big(2\delta_{18}-\frac{\alpha_5}{\alpha_3}\beta_{20}\Big),\nonumber
\end{align}
\begin{align}
&\eta_{10}=\frac{8\alpha_2}{3}\Big(\frac{\beta_{20}}{\alpha_3}-\frac{\delta_{18}}{\alpha_5}\Big),\hspace{2mm}\eta_{11}=\frac{-2\beta_{17}}{3},\hspace{2mm}\eta_{12}=\frac{-2\beta_{18}}{3},\hspace{2mm}\eta_{13}=\frac{2\beta_{19}}{3},\nonumber\\
&\eta_{14}=\frac{\alpha_4}{108\alpha_1}\Big[6\delta_1+\frac{\alpha_4}{\alpha_1}(\beta_1-7\beta_2)\Big], \hspace{2mm}\eta_{15}=\frac{\alpha_4\alpha_1}{54\alpha_1^2}(\beta_1-7\beta_2)+\frac{\alpha_2\delta_1}{\alpha_118}-\frac{2\delta_{15}}{3},\nonumber\\
&\eta_{16}=\frac{1}{3}\Big[\frac{\alpha_2}{\alpha_1}(\delta_2+\frac{2\alpha_4\beta_2}{\alpha_1})-2\delta_{22}\Big],\hspace{2mm}\eta_{17}=\frac{\alpha_4}{3\alpha_1}\Big(\delta_2+\frac{\alpha_4}{\alpha_1}\beta_2\Big),\hspace{2mm}\eta_{18}=\frac{\alpha_{5}}{12\alpha_3}\Big(2\delta_{18}-\frac{\alpha_5\beta_{20}}{\alpha_3}\Big)\nonumber\\
&\eta_{19}=\frac{2\delta_3}{3},\hspace{2mm}\eta_{20}=\frac{2\delta_4}{3},\hspace{2mm}\eta_{21}=\frac{-2\delta_{17}}{3},\hspace{2mm}\eta_{22}=\frac{-2\delta_{19}}{3}.
\end{align}
  
With all these preliminaries, equation (\ref{Na4f 5D}) is the most general 5D action of gravity coupled with U(1) gauge and SU(2) Yang-Mills fields that contains all the 4-derivative interaction terms. 

Our next step is to truncate the 5D action to 2D using the following ansatz
\begin{align}
      &ds^2_{(5)}=ds^2_{(2)}+\phi(t,z)^{\frac{2}{3}}(dx^2+dy^2+dz^2),\label{AMA}\\
      &A_Mdx^M=A_{\mu}dx^{\mu},\hspace{2mm} A_{\mu}\equiv A_{\mu}(x^{\nu}),\label{AAA}\\
       &A_M^{(a)}dx^M=A_{\mu}^{(a)}dx^{\mu},\hspace{2mm} A_{\mu}^{(a)}\equiv A_{\mu}^{(a)}(x^{\nu}).\label{ANA}
\end{align}
Using the above ansatz (\ref{AMA})-(\ref{ANA}) in (\ref{Na4f 5D}) we finally obtain
\begin{align}\label{NA 2D}
    S_{(2D)} =& \int d^2x\sqrt{-g_{(2)}}\phi\Bigg[\eta_1\lambda+\eta_2R-\frac{\eta_3}{4}F^2-\frac{\eta_4}{4} \big(F^{(a)}\big )^2+\eta_6\Big[(R_{\mu\nu\alpha\beta})^2+\frac{3}{4}\big(\bigtriangledown_{\mu}\phi^{\frac{2}{3}}\big)^4\nonumber\\
&+4\Big\{\frac{3}{4}(\bigtriangledown_{\lambda}\phi^{\frac{2}{3}})(\bigtriangledown_{\beta}\phi^{\frac{2}{3}})\Gamma^{\lambda}_{\alpha\mu}\Gamma^{\beta}_{\rho\sigma}g^{\alpha\rho}g^{\mu\sigma}+\frac{2}{3}\Gamma^{\lambda}_{\alpha\mu}(\bigtriangledown_{\lambda}\phi^{\frac{2}{3}})(\bigtriangledown^{\alpha}\phi)(\bigtriangledown^{\mu}\phi)\phi^{\frac{-4}{3}}-\nonumber\\
&\Gamma^{\lambda}_{\alpha\mu}(\bigtriangledown_{\lambda}\phi^{\frac{2}{3}})\{\partial_{\beta}(\bigtriangledown_{\sigma}\phi)\}g^{\alpha\beta}g^{\mu\sigma}\phi^{\frac{-1}{3}}+\frac{4}{27}(\bigtriangledown_{\mu}\phi)^4\phi^{\frac{-8}{3}}-\frac{4}{9}\phi^{\frac{-5}{3}}(\bigtriangledown^{\alpha}\phi)(\bigtriangledown^{\mu}\phi)\{\partial_{\alpha}(\bigtriangledown_{\mu}\phi)\}\nonumber\\
&-\frac{1}{3}\{\partial_{\alpha}(\bigtriangledown_{\mu}\phi)\}\{\partial_{\beta}(\bigtriangledown_{\rho}\phi)\}g^{\alpha\beta}g^{\mu\rho}\phi^{\frac{-2}{3}}\Big\}\Big]+\eta_7F^4+\eta_8F^{\mu\nu}F_{\nu\lambda}F^{\lambda\sigma}F_{\sigma\mu}+\nonumber\\
&\eta_9\bigtriangledown_{\mu}F^{\mu\nu}\bigtriangledown^{\lambda}F_{\lambda\nu}+\eta_{14}\big(\big(F^{(a)}\big)^2\big)^2+\eta_{15}F^2\big(F^{(a)}\big)^2+\eta_{16}F^{(a)\mu\nu}F^{(a)}_{\lambda\nu}F^{\lambda\sigma}F_{\mu\sigma}+\nonumber\\
&\eta_{17}F^{(a)\mu\nu}F^{(a)}_{\lambda\nu}F^{(b)\lambda\sigma}F^{(b)}_{\mu\sigma}+\eta_{19}R_{\mu\nu\sigma\lambda}F^{(a)\mu\nu}F^{(a)\sigma\lambda}+\eta_{20}R_{\mu\nu\lambda\sigma}F^{(a)\mu\lambda}F^{(a)\nu\sigma}+\nonumber\\
&\eta_{21}F_{\mu\nu}F_{\lambda\sigma}F^{(a)\mu\nu}F^{(a)\lambda\sigma}\Bigg],
\end{align}
where we identify  (\ref{NA 2D}) as the most general 2D action of gravity coupled with U(1) gauge and SU(2) Yang-Mills fields that contains all the 4-derivative interaction terms.
\section{Most general 4-derivative action in 5D}\label{gen action 5D}
The purpose of this Appendix is to discuss the most general 5D action of gravity coupled with U(1) gauge fields that contains the 4-derivative interaction terms\footnote{This can be achieved by simply removing the SU(2) Yang-Mills fields in Appendix \ref{Non-abelian}.}. 

In the first place, we consider the most general 2-derivative action of the following form
\begin{eqnarray}\label{A2d}
S^{(2)}=&& \int d^5x\sqrt{-g}\Bigg[12 + R-\frac{\alpha_1}{4}F^2+\frac{\alpha_2}{3}\epsilon^{MNOPQ}A_MF_{NO}F_{PQ}\Bigg],
\end{eqnarray}
where $\alpha_1$ and $\alpha_2$ are the respective coupling constants.

Next, we note down the most general 4-derivative gauge invariant interaction terms as follows
\begin{eqnarray}\label{A4d}
S^{(4)}&=&\int d^5x\sqrt{-g} \Big [\beta_1 R^2+\beta_2 [R_{MN}]^2 + \beta_3 [R_{MNOP}]^2+\beta_4 RF^2+ \beta_5R^{MN}F_{MO}{F_{N}}^O +\nonumber\\
&&\beta_6 R_{MNOP}F^{MN}F^{OP}+\beta_7 R_{MNOP}F^{MO}F^{NP}+\beta_8 F^4+ \beta_9 F^{MN}F_{NO}F^{OP}F_{PM}+\nonumber\\
&&\beta_{10}\bigtriangledown^MF_{MN}\bigtriangledown^O{F_{O}}^N+\beta_{11}\bigtriangledown_MF_{NO}\bigtriangledown^MF^{NO}+\beta_{12}\bigtriangledown_MF_{NO}\bigtriangledown^NF^{MO}+\nonumber\\
&&\beta_{13}\bigtriangledown^2F^2+\beta_{14}\bigtriangledown_M\bigtriangledown^NF_{NO}F^{MO}+\beta_{15}\bigtriangledown^N\bigtriangledown_MF_{NO}F^{MO}+\nonumber\\
&&\epsilon^{MNOPQ}\big\{F_{MN}(\beta_{16}F_{OP}\bigtriangledown^RF_{RQ}+\beta_{17}F_{OR}\bigtriangledown^RF_{PQ}+\beta_{18}F_{OR}\bigtriangledown_PF_{QS}g^{RS})+\nonumber\\
&&\beta_{19}A_M R_{NOKL}{R_{PQ}}^{KL}\big\}+\beta_{20}\epsilon_{NOPQR}\epsilon^{NIJKL}F^{OP}F^{QR}F_{IJ}F_{KL}\Big],
\end{eqnarray}
where $\beta_i$ $(i=1,..,20)$ are the respective coupling constants. On adding (\ref{A4d}) in (\ref{A2d}), we obtain the most general action for gravity (coupled to U(1) gauge fields) that contains 4-derivative interaction terms. Moreover, it is also possible to eliminate the various interaction terms in (\ref{A4d}) using a proper redefinition of fields as discussed below.

Consider the following redefinition of fields  $$g^{RS}\rightarrow g^{RS}+\delta g^{RS},\hspace{2mm}A_N\rightarrow A_N+\delta A_N$$ such that the action transform as $S\rightarrow S'=S+\delta S$. 

The most general 2-derivative variation of fields are given by
\begin{eqnarray}
\delta g^{RS} &=& \mu_1R^{RS}+\mu_2F^{RP}{F^S}_P+\mu_3Rg^{RS}+\mu_4F^2g^{RS}+\mu_5g^{RS}\label{aa1}\\
\delta A_N&=&\lambda_1A_N+\lambda_2\bigtriangledown^MF_{MN}+\lambda_3\epsilon_{NOPQR}F^{OP}F^{QR}.\label{aa2}
\end{eqnarray}
where $\mu_i$ and $\lambda_j$ ($i=1,..,4$ and $j=1,..,3$) are the respective coupling constants. 

 In order to get rid of various interaction terms in (\ref{A2d}) and (\ref{A4d}), we transformed the action $S$ (where $S=S_{5D}^{(2)}+S_{5D}^{(4)}$) into $S'$ using (\ref{aa1}) and (\ref{aa2}) with the following particular choice of constants 
\begin{eqnarray}
&&\mu_1=-\frac{4\beta_2}{3},\hspace{3mm}\mu_2=-\frac{2}{3}\alpha_1\beta_2 ,\hspace{3mm}\mu_3=\frac{4}{9}(2\beta_1+\beta_2), \hspace{3mm}\mu_4=\frac{2\alpha_1}{27}(2\beta_2+\beta_1),\nonumber\\
&&\mu_5 = -\frac{2}{3},\hspace{3mm}\lambda_1=\frac{-1}{3},\hspace{3mm}\lambda_2=\frac{4\alpha_1}{3\alpha_2^2}\beta_{20},\hspace{3mm}\lambda_3=\frac{4\beta_{20}}{3\alpha_2},
\end{eqnarray}
which yields
\begin{align}\label{FA4d}
    S'=&\int d^5x\sqrt{-g}\Big [(12+R)-\frac{\eta_1}{4}F^2+\eta_2[R_{MNOP}]^2+\eta_3 F^4+\eta_4F^{SP}F_{PR}F^{RQ}F_{QS}+ \nonumber\\
&\eta_5\bigtriangledown_MF^{MN}\bigtriangledown^OF_{ON}+\epsilon^{MNOPQ}(\eta_{6}F_{MN}F_{OP}\bigtriangledown^RF_{RQ}+\eta_{7}F_{MN}F_{OR}\bigtriangledown^RF_{PQ}+\nonumber\\
&\eta_{8}F_{MN}F_{OR}\bigtriangledown_P{F_Q}^R+\eta_{9}A_MR_{NOIJ}{R_{PQ}}^{IJ}) \Big ]
\end{align}
where $\eta_i$ ($i=1,..,9$) are the new coupling constants respectively. 

We can express the new coupling constants, $\eta_i$ in terms of old coupling constants, $\alpha_j$ and $\beta_k$ as follows
\begin{align}
    &\eta_1=\frac{\alpha_1}{6}(4\beta_2+20\beta_1)+\frac{\alpha_1}{4},\hspace{2mm}\eta_2=\frac{\beta_3}{2},\hspace{2mm}\eta_3=\frac{\alpha_1^2}{288}(\beta_1-7\beta_2)-\frac{\beta_8}{2},\hspace{2mm}\eta_4=\frac{1}{8}(\alpha_1^2\beta_2-4\beta_9),\nonumber\\ &\eta_5=\frac{\alpha_1^2}{2\alpha_2^2}\beta_{20},\hspace{2mm}
    \eta_{6}=\alpha_1\Big(\frac{\beta_{20}}{\alpha_2}\Big),\hspace{3mm}\eta_{7}=\frac{-\beta_{17}}{4},\hspace{3mm}\eta_{8}=\frac{-\beta_{18}}{4},\hspace{3mm}\eta_{9}=\frac{\beta_{19}}{4}.
\end{align}

Notice that, we have taken out a common factor $\frac{8}{3}$ in (\ref{FA4d}) and impose the condition on $\beta_{1,2}$ such that $5\beta_1+\beta_2=-\frac{1}{8}$. With all these preliminaries, we end up with a most general 5D action (\ref{FA4d}) of gravity coupled with U(1) gauge fields that contains all the 4-derivative interaction terms. 

\section{Covariance of the 2D action (\ref{action 2D})}\label{covariance}
In this section, we will discuss the general covariance of the action (\ref{action 2D}). After a careful observation,  we realize that the $\Gamma^{\alpha}_{\beta\lambda}$s appear in such a combination that the entire expression transform as a ``scalar'' under general coordinate transformation (GCT). 

As an illustration, let's look at a particular term in (\ref{action 2D}) as mentioned below:
\begin{align}
\Gamma^{\sigma}_{\alpha\mu}(\bigtriangledown_{\sigma}\phi)(\bigtriangledown^{\alpha}\phi)(\bigtriangledown^{\mu}\phi)=\Gamma^{\sigma}_{\alpha\mu}A_{\sigma}B^{\alpha}C^{\mu},\label{C1}
\end{align}
where $A_{\sigma},B^{\alpha}$ and $C^{\mu}$ are the tensors of rank 1. 

A straightforward computation reveals that under GCT 
\begin{eqnarray}\label{C3}
\Gamma^{\sigma}_{\alpha\mu}A_{\sigma}B^{\alpha}C^{\mu}\rightarrow\Gamma^{'\sigma}_{\alpha\mu}A'_{\sigma}B^{'\alpha}C^{'\mu}
&=&\frac{\partial x^{'\sigma}}{\partial x^{\beta}}\frac{\partial x^{\rho}}{\partial x^{'\mu}}\frac{\partial x^{\tau}}{\partial x^{'\alpha}}\Gamma^{\beta}_{\tau\rho}\frac{\partial x^{\tilde{\beta}}}{\partial x^{'\sigma}}\frac{\partial x^{'\alpha}}{\partial x^{\tilde{\tau}}}\frac{\partial x^{'\mu}}{\partial x^{\tilde{\rho}}}A_{\tilde{\beta}}B^{\tilde{\tau}}C^{\tilde{\rho}}\nonumber\\
&&-\frac{\partial x^{\tilde{\tau}}}{\partial x^{'\alpha}}\frac{\partial^2x^{'\sigma}}{\partial x^{\tilde{\tau}}\partial x^{\tilde{\beta}}}\frac{\partial x^{\tilde{\beta}}}{\partial x^{'\mu}}\frac{\partial x^{\beta}}{\partial x^{'\sigma}}\frac{\partial x^{'\alpha}}{\partial x^{\tau}}\frac{\partial x^{'\mu}}{\partial x^{\rho}}A_{\beta}B^{\tau}C^{\rho} \nonumber\\\nonumber 
&=&\Gamma^{\beta}_{\tau\rho}A_{\tilde{\beta}}B^{\tilde{\tau}}C^{\tilde{\rho}}\delta^{\tilde{\beta}}_{\beta}\delta^{\rho}_{\tilde{\rho}}\delta^{\tau}_{\tilde{\tau}}-\delta^{\tilde{\tau}}_{\tau}\delta^{\tilde{\beta}}_{\rho}\frac{\partial}{\partial x^{\tilde{\tau}}}\Big(\frac{\partial x^{'\sigma}}{\partial x_{\tilde{\beta}}}\Big)\frac{\partial x^{\beta}}{\partial x^{'\sigma}}A_{\beta}B^{\tau}C^{\rho}\nonumber\\
&=&\Gamma^{\beta}_{\tau\rho}A_{\beta}B^{\tau}C^{\rho}-\frac{\partial}{\partial x^{\tau}}\Big(\frac{\partial x^{'\sigma}}{\partial x_{\tilde{\beta}}}\Big)\frac{\partial x^{\beta}}{\partial x^{'\sigma}}A_{\beta}B^{\tau}C^{\tilde{\beta}}.
\end{eqnarray}

 Notice that, the last term in (\ref{C3}) vanishes identically as shown below
\begin{eqnarray}\label{C4}
\Bigg[\frac{\partial}{\partial x^{\tau}}\Big(\frac{\partial x^{'\sigma}}{\partial x_{\tilde{\beta}}}\Big)\frac{\partial x^{\beta}}{\partial x^{'\sigma}}\Bigg]A_{\beta}B^{\tau}C^{\tilde{\beta}}&=&\Bigg[\frac{\partial}{\partial x^{\tau}}\Big(\frac{\partial x^{'\sigma}}{\partial x_{\tilde{\beta}}}\frac{\partial x^{\beta}}{\partial x^{'\sigma}}\Big)-\frac{\partial x^{'\sigma}}{\partial x_{\tilde{\beta}}}\frac{\partial^2x^{\beta}}{\partial x^{\tau}\partial x^{'\sigma}}\Bigg]A_{\beta}B^{\tau}C^{\tilde{\beta}}\nonumber\\
&=&\Bigg[\frac{\partial}{\partial x^{\tau}}\Big(\delta^{\beta}_{\tilde{\beta}}\Big)-\frac{\partial x^{'\sigma}}{\partial x_{\tilde{\beta}}}\frac{\partial}{\partial x^{'\sigma}}\Big(\frac{\partial x^{\beta}}{\partial x^{\tau}}\Big)\Bigg]A_{\beta}B^{\tau}C^{\tilde{\beta}}\nonumber\\
&=& 0.
\end{eqnarray}

 \section{Solving the constants \texorpdfstring{$D_{i}$s}{Di}}\label{const}
 In order to cure the divergences in the Gibbons-Hawking-York term and Einstein-Hilbert action (\ref{SGHY})-(\ref{SEH}), we set the constants $D_{i}$s (\ref{SCT}) such that the coefficients of each divergent term vanish identically. 
 
 Below, we note  down the coefficients of each divergent terms as
 \begin{align}
     &\text{coefficient of $\frac{1}{z^\frac{10}{3}}$}\hspace{2mm}:\hspace{2mm}\frac{-2032}{45}-\frac{103172}{585}D_1-\frac{14188}{195}\sqrt{\frac{2}{3}}D_2+\frac{11104}{45}D_3=0,\label{CONST1}\\
     &\text{coefficient of $\frac{1}{z^\frac{11}{3}}$}\hspace{2mm}:\hspace{2mm}-\frac{24}{7}D_1-\frac{4}{7}\sqrt{6}D_2=0,\label{CONST2}\\
     &\text{coefficient of $\frac{1}{z^\frac{14}{3}}$}\hspace{2mm}:\hspace{2mm}\frac{64}{21}+\frac{18}{7}D_1+\frac{4}{7}\sqrt{\frac{2}{3}}D_2-\frac{32}{7}D_3=0,\label{CONST3}\\
      &\text{coefficient of $\xi\log(z)$}\hspace{2mm}:\hspace{2mm}D_4+\frac{11213}{3123}=0,\label{CONST4}\\
            &\text{coefficient of $\kappa\log(z)$}\hspace{2mm}:\hspace{2mm}12D_4C_3+\frac{19868C_3}{347}+\frac{D_4}{6}C_9+\frac{1844}{9369}C_9=0,\label{CONST5}\\
            &\text{coefficient of $\frac{1}{z^2}$}\hspace{2mm}:\hspace{2mm}\frac{16112}{1041}\kappa-\frac{871}{6246}=0,\label{CONST6}\\
            &\text{coefficient of $\frac{\xi}{z^3}$}\hspace{2mm}:\hspace{2mm}D_5+\frac{871}{3123}=0,\label{CONST7}\\
            &\text{coefficient of $\frac{\kappa}{z^3}$}\hspace{2mm}:\hspace{2mm}D_6+\frac{871}{3123}=0.\label{CONST8}
 \end{align}
 
 On solving (\ref{CONST1})-(\ref{CONST8}), we get the following values as solutions
 \begin{align}
    &D_1=-2.5795,\hspace{1mm}D_2=6.3186,\hspace{1mm}D_3=-0.1394,\hspace{1mm}D_4=-3.5904,\hspace{1mm}D_5=D_6=-0.2788,\hspace{1mm}\nonumber\\ &C_3=0.0283C_9,\hspace{1mm}\kappa=0.0090.
\end{align}

\section{Cardy formula}\label{cardy}
Cardy formula \cite{Cardy:1986ie} measures the degrees of freedom (and hence the entropy) of a two-dimensional conformal field theory (CFT$_{2}$)\footnote{For the  generalization to d-dimensional CFT, see \cite{Verlinde:2000wg}} using the central charge (c) and the conformal weight for the ground state ($\Delta$). In the present Section, we derive the expression for the Cardy formula (\ref{s1}) that we have used in Section (\ref{ccs1}). We start by computing the partition function of CFT$_{2}$ on torus (T$^{2}$) and then use this partition function to estimate the entropy. 

The partition function of CFT$_{2}$ on torus (T$^{2}$) of modular parameter $\tau=x^1+ix^0$ is given by \cite{Cardy:1986ie}
\begin{align}\label{cd1}
    Z=\text{Tr}\Big[e^{-2\pi(Im\tau)H}e^{i2\pi(Re\tau)P}\Big],
\end{align}
where $H$ and $P$ are the time and space translation on the cylinder respectively and the factor of $2\pi$ in the definition of $Z$ is merely a convention.

One can derive $H$ and $P$ from the stress tensor $T_{\mu\nu}$ as 
\begin{align}\label{cd2}
    H=\frac{1}{2\pi}\int dx^1T_{00}\hspace{2mm}\text{and}\hspace{2mm}  P=\frac{1}{2\pi}\int dx^1T_{01},
\end{align}
where $T_{00}$ and $T_{01}$ can be expressed in terms of the stress tensor on the cylinder ($T_{cyl}$) as 
\begin{align}\label{cd3}
    T_{00}=-\big(T_{cyl}(z)+T_{cyl}(\overline{z})\big)\hspace{2mm}\text{and}\hspace{2mm}T_{01}=-\big(T_{cyl}(z)-T_{cyl}(\overline{z})\big).
\end{align}

The stress tensor on the cylinder, $ T_{cyl}(z)$ is given by \cite{Tong:2009np}-\cite{icts}
\begin{align}\label{cd4}
    T_{cyl}(z)=-\sum e^{-inz}L_n+\frac{c}{24},
\end{align}
where $L_n$ are the virasoro generators that satisfy the following commutation relations
\begin{align}\label{cd5}
    &[L_n,L_m]=(n-m)+\frac{c}{12}n(n^2-1)\delta_{n+m,0}\\
     &[\overline{L}_n,\overline{L}_m]=(n-m)+\frac{c}{12}n(n^2-1)\delta_{n+m,0}\\
     &[L_n,\overline{L}_m]=0.
\end{align}

Using equations (\ref{cd2})-(\ref{cd4}) in (\ref{cd1}), we get\footnote{Since central charge is the real number, therefore we use $\overline{c}=c$ in (\ref{cd6}).}
\begin{align}\label{cd6}
    Z(\tau,\overline{\tau})=\text{Tr}\Big[e^{2\pi i\tau\big(L_0-\frac{c}{24}\big)}e^{2\pi i\overline{\tau}\big(\overline{L}_0-\frac{c}{24}\big)}\Big].
\end{align}

Next, we trace (\ref{cd6}) over the energy eigen state $|E\rangle$ that satisfy the following eigen value equation
\begin{align}
    L_0|E\rangle=\Delta|E\rangle,
\end{align}
which yields\footnote{For simplicity, we have consider only first part of the trace.}
\begin{align}\label{cd7}
    Z(\tau)=\int_0^{\infty}d\Delta\rho(\Delta)e^{2\pi i\tau\big(\Delta-\frac{c}{24}\big)},
    \end{align}
where $\rho(\Delta)$ is the density of states correspond to the energy $\Delta$.

One can evaluate $\rho(\Delta)$ by taking the inverse Laplace transformation of (\ref{cd7}) as follows
\begin{align}\label{cd8}
    \rho(\Delta)=\oint_c d\tau Z(\tau)e^{-2\pi i\tau\big(\Delta-\frac{c}{24}\big)}.
\end{align}

We are interested in computing the entropy of CFT$_2$ in the high temperature limit (or $\Delta>>1$). Therefore, in this limit, the integral (\ref{cd8}) is dominated by $Z(\tau\rightarrow0)$. In order to find $Z(\tau)$ in the limit $\tau\rightarrow0$, we utilize the important fact that the partition function (\ref{cd6}) is modular invariant i.e. $Z(\tau)=Z(-\frac{1}{\tau})$. This can be understood in terms of the geometry of torus as described below. 

Consider a torus (T$^2$) which is parametrize by the two coordinates i.e. $\sigma_1$ and $\sigma_2$ in the range $0\leq\sigma_1\leq2\pi$ and $0\leq\sigma_2\leq2\pi$. The general metric of the torus is given by
\begin{align}\label{t1}
   ds^2=|d\sigma_1+\tau d\sigma_2|^2 \hspace{2mm},
\end{align}
where $\tau\in\mathbb{C}$ is the modular parameter. It is easy to check that (\ref{t1}) is invariant (up to conformal factor) under the following $SL(2,\mathbb{Z})$ transformation  
\begin{align}
\begin{pmatrix}
 \sigma_1\\
 \sigma_2
\end{pmatrix}=\begin{pmatrix}
 d & b\\
 c & a
\end{pmatrix}\begin{pmatrix}
 \sigma'_1\\
 \sigma'_2
\end{pmatrix}\hspace{2mm},\hspace{2mm}\tau'=\frac{a\tau+b}{c\tau+d}\hspace{2mm},
\end{align}
where $a,b,c,d\in\mathbb{Z}$ and $ad-bc=1$.

Therefore we conclude that the modular parameter $\tau'$ is equivalent to $\tau$ and for a particular choice of constants i.e. $a=0,b=-1,c=1,d=0$, we get $\tau'=-\frac{1}{\tau}$. 

Using this property in (\ref{cd6}), we obtain 
\begin{align}\label{cd9}
    Z(\tau)=e^{2\pi i\frac{1}{\tau}\frac{c}{24}}\overline{Z}(-\frac{1}{\tau}),\hspace{2mm}\text{where}\hspace{2mm}\overline{Z}(-\frac{1}{\tau})\equiv \text{Tr}e^{-2\pi i \frac{1}{\tau}L_0}.
\end{align}

In the limit $\tau\rightarrow 0$, the dominant contribution in $\overline{Z}(-\frac{1}{\tau})$ comes from the lowest eigen value of the operator $L_0$ which is set to be zero without any loss of generality. Therefore, (\ref{cd9}) reduces to $ Z(\tau)=e^{2\pi i\frac{1}{\tau}\frac{c}{24}}$. 

Using this expression in (\ref{cd8}), we finally obtain 
\begin{align}\label{cd10}
    \rho(\Delta)=\oint_c d\tau e^{-2\pi i \big(\tau\big(\Delta-\frac{c}{24}\big)-\frac{1}{\tau}\frac{c}{24}\big)}.
\end{align}

One can approximate the above integral (\ref{cd10}) using the saddle point approximation as 
\begin{align}\label{cd11}
    \rho(\Delta)\approx e^{-2\pi i f(\tau*)},\hspace{2mm}\text{where}\hspace{2mm}f(\tau)=\tau\big(\Delta-\frac{c}{24}\big)-\frac{1}{\tau}\frac{c}{24}
\end{align}
 and $\tau*$ is calculated using the following condition
 \begin{align}\label{cd12}
     \frac{d f(\tau)}{d \tau}\Big|_{\tau=\tau*}=0,\hspace{2mm}\text{which yields}\hspace{2mm}\tau*=i\sqrt{\frac{c}{24\big(\Delta-\frac{c}{24}\big)}}.
 \end{align}
 
On plugging the above value of $\tau*$ (\ref{cd12}) in (\ref{cd11}), we obtain the density of states $\rho(\Delta)$ in the limit $\Delta>>1$. 

The Cardy formula for the black hole is defined in terms of density of states $\rho(\Delta)$ \cite{icts} as 
 \begin{align}\label{CDF}
     S_{Cardy}=\log[\rho(\Delta)].
 \end{align}
 Using (\ref{cd12}) and (\ref{cd11}) in (\ref{CDF}), we obtain the Cardy formula for a 2D black hole in the limit $\Delta>>1$ as
  \begin{align}\label{CDF1}
    S_{Cardy}=2\pi\sqrt{\frac{c\Delta}{6}}.
      \end{align}
 Equation (\ref{CDF1}) is the entropy of a two-dimensional CFT.
\section{Properties of the potential \texorpdfstring{$V(\psi)$}{Vpsi}}\label{vp}
In this Appendix, we discuss the stability of the potential function $V(\psi)$ for the generalised Liouville theory as constructed in (\ref{NHT8}). Finally, we generalise this theory (\ref{NHT8}) by considering 4-derivative interactions.

In order to study the stability of the potential, we first note down the extrema of $V(\psi)$ (\ref{NHT8}) by setting
\begin{align}\label{fa3}
    \frac{dV(\psi)}{d\psi}\Bigg|_{\psi=\psi_{i}}=0,\hspace{2mm}(i=1,2,3)
\end{align}
which reveals
\begin{align}
    \psi_1&=-\frac{1}{2}\sqrt{\frac{c_H}{3\pi}}\label{fa4},\\
   \psi_2&= \frac{1}{2}\sqrt{\frac{c_H}{3\pi}}\text{Productlog}\Bigg[-\pi\sqrt{\frac{3b^2\xi}{2c_H^2}}\Bigg]\label{fa5},\\
    \psi_3&= \frac{1}{2}\sqrt{\frac{c_H}{3\pi}}\text{Productlog}\Bigg[\pi\sqrt{\frac{3b^2\xi}{2c_H^2}}\Bigg]\label{fa6},
\end{align}
where we express $\Phi_H$ in terms of the central charge $c_H$ (\ref{cch}) associated with the theory (\ref{NHT8}).

Using (\ref{fa4})-(\ref{fa6}), we find that the potential $V(\psi)$ exhibits local minima at $\psi_1$ and $\psi_3$ (if $2c_H^2>3\xi\pi^2 b^2 e$) which we identify as the possible vacuua of the theory (\ref{NHT8}). However, notice that in the limit of large central charge $\psi_1\rightarrow-\infty$ which therefore corresponds to the most stable vacuua of the theory (\ref{NHT8}).

Finally, we write down the most general action  (\ref{NHT8}) by considering 4-derivative interactions. Under the following field redefinition
\begin{align}
    \phi=q\Phi_H\psi\hspace{2mm},\hspace{2mm}g_{\mu\nu}\rightarrow e^{2\sigma}g_{\mu\nu}\hspace{2mm},\text{where}\hspace{2mm}\sigma=\frac{\psi}{q\Phi_H},
\end{align}
4-derivative interactions in (\ref{action 2D}) transform as \begin{align}
    S^{(4)}=&q\Phi_H\kappa\int d^2x\sqrt{-g}\psi\Bigg[e^{-2\sigma}\Big\{\big[R_{\mu\nu\alpha\beta}-\bigtriangledown^2\sigma(g_{\mu\alpha}g_{\nu\beta}-g_{\mu\beta}g_{\nu\alpha})\big]\big[R^{\mu\nu\alpha\beta}-\bigtriangledown^2\sigma(g^{\mu\alpha}g^{\nu\beta}\nonumber\\
    &-g^{\mu\beta}g^{\nu\alpha})\big]\Big\}+\frac{3}{4}(\bigtriangledown_{\mu}(q\Phi_H\psi)^{\frac{2}{3}})^4e^{-2\sigma}+4e^{-2\sigma}\Big\{\frac{3}{4}(\bigtriangledown_{\lambda}(q\Phi_H\psi)^{\frac{2}{3}})(\bigtriangledown_{\beta}(q\Phi_H\psi)^{\frac{2}{3}})(\Gamma^{\lambda}_{\alpha\mu}+\nonumber\\
    &\frac{e^{-2\sigma}}{2}(\delta^{\lambda}_{\alpha}\bigtriangledown_{\mu}e^{2\sigma}+\delta^{\lambda}_{\mu}\bigtriangledown_{\alpha}e^{2\sigma}-g_{\alpha\mu}g^{\lambda\bar{\sigma}}\bigtriangledown_{\bar{\sigma}}e^{2\sigma}))(\Gamma^{\beta}_{\rho\bar{\sigma}}+\frac{e^{-2\sigma}}{2}(\delta^{\beta}_{\rho}\bigtriangledown_{\bar{\sigma}}e^{2\sigma}+\delta^{\beta}_{\bar{\sigma}}\bigtriangledown_{\rho}e^{2\sigma}-\nonumber\\
    &g_{\rho\bar{\sigma}}g^{\beta\lambda}\bigtriangledown_{\lambda}e^{2\sigma}))g^{\alpha\rho}g^{\mu\bar{\sigma}}+\frac{2}{3}    (\Gamma^{\lambda}_{\alpha\mu}+\frac{e^{-2\sigma}}{2}(\delta^{\lambda}_{\alpha}\bigtriangledown_{\mu}e^{2\sigma}+\delta^{\lambda}_{\mu}\bigtriangledown_{\alpha}e^{2\sigma}-g_{\alpha\mu}g^{\lambda\bar{\sigma}}\bigtriangledown_{\bar{\sigma}}e^{2\sigma}))\times\nonumber\\
    &\Big(\bigtriangledown_{\lambda}(q\Phi_H\psi)^{\frac{2}{3}}\Big)(\bigtriangledown^{\alpha}(q\Phi_H\psi))(\bigtriangledown^{\mu}(q\Phi_H\psi))(q\Phi_H\psi)^{\frac{-4}{3}}-(\Gamma^{\lambda}_{\alpha\mu}+\frac{e^{-2\sigma}}{2}(\delta^{\lambda}_{\alpha}\bigtriangledown_{\mu}e^{2\sigma}+\nonumber\\
    &\delta^{\lambda}_{\mu}\bigtriangledown_{\alpha}e^{2\sigma}-g_{\alpha\mu}g^{\lambda\bar{\sigma}}\bigtriangledown_{\bar{\sigma}}e^{2\sigma}))\Big(\bigtriangledown_{\lambda}(q\Phi_H\psi)^{\frac{2}{3}}\Big)\{\partial_{\beta}(\bigtriangledown_{\tilde{\sigma}}(q\Phi_H\psi))\}g^{\alpha\beta}g^{\mu\tilde{\sigma}}(q\Phi_H\psi)^{\frac{-1}{3}}-\nonumber\\
    &\frac{4}{9}(q\Phi_H\psi)^{\frac{-5}{3}}(\bigtriangledown^{\alpha}(q\Phi_H\psi))(\bigtriangledown^{\mu}(q\Phi_H\psi))\{\partial_{\alpha}(\bigtriangledown_{\mu}(q\Phi_H\psi))\}+\frac{4}{27}(\bigtriangledown_{\mu}(q\Phi_H\psi))^4\times\nonumber\\
    &(q\Phi_H\psi)^{\frac{-8}{3}}-\frac{1}{3}\{\partial_{\alpha}(\bigtriangledown_{\mu}(q\Phi_H\psi))\}\{\partial_{\beta}(\bigtriangledown_{\rho}(q\Phi_H\psi))\}g^{\alpha\beta}g^{\mu\rho}(q\Phi_H\psi)^{\frac{-2}{3}}\Big\}+e^{-6\sigma}F^4+\nonumber\\
    &e^{-6\sigma}F^{\mu\nu}F_{\nu\lambda}F^{\lambda\bar{\sigma}}F_{\bar{\sigma}\mu}+e^{-4\sigma}g^{\mu\Bar{\mu}}g^{\nu\Bar{\nu}}g^{\lambda\Bar{\lambda}}(\bigtriangledown_{\mu}F_{\Bar{\mu}\Bar{\nu}}-\frac{e^{-2\sigma}}{2}(\delta^{\rho}_{\mu}\bigtriangledown_{\Bar{\mu}}e^{2\sigma}+\delta^{\rho}_{\bar{\mu}}\bigtriangledown_{\mu}e^{2\sigma}-\nonumber\\
    &g_{\mu\bar{\mu}}g^{\rho\bar{\sigma}}\bigtriangledown_{\bar{\sigma}}e^{2\sigma})F_{\rho\bar{\nu}}-\frac{e^{-2\sigma}}{2}(\delta^{\rho}_{\mu}\bigtriangledown_{\Bar{\nu}}e^{2\sigma}+\delta^{\rho}_{\bar{\nu}}\bigtriangledown_{\mu}e^{2\sigma}-g_{\mu\bar{\nu}}g^{\rho\bar{\sigma}}\bigtriangledown_{\bar{\sigma}}e^{2\sigma})F_{\bar{\mu}\rho})\times(\bigtriangledown_{\bar{\lambda}}F_{\lambda\nu}-\nonumber\\
    &\frac{e^{-2\sigma}}{2}(\delta^{\rho}_{\lambda}\bigtriangledown_{\Bar{\lambda}}e^{2\sigma}+\delta^{\rho}_{\bar{\lambda}}\bigtriangledown_{\lambda}e^{2\sigma}-g_{\lambda\bar{\lambda}}g^{\rho\bar{\sigma}}\bigtriangledown_{\bar{\sigma}}e^{2\sigma})F_{\rho\nu}-\frac{e^{-2\sigma}}{2}(\delta^{\rho}_{\nu}\bigtriangledown_{\Bar{\lambda}}e^{2\sigma}+\delta^{\rho}_{\bar{\lambda}}\bigtriangledown_{\nu}e^{2\sigma}-\nonumber\\
    &g_{\nu\bar{\lambda}}g^{\rho\bar{\sigma}}\bigtriangledown_{\bar{\sigma}}e^{2\sigma})F_{\lambda\rho})\Bigg]\label{4v}.
\end{align}

Notice that, in arriving at (\ref{4v}), we express the Riemann tensor in 2D as $$R_{\mu\nu\alpha\beta}=\frac{R}{2}(g_{\mu\alpha}g_{\nu\beta}-g_{\mu\beta}g_{\nu\alpha}),$$ where $R$ is the Ricci scalar in 2D. Equation (\ref{4v}) represents corrections to the Liouville theory (\ref{NHT8}) due to the presence of 4-derivative interactions in (\ref{action 2D}).

\end{document}